\documentclass[11pt]{article}
\usepackage[english]{babel}
\usepackage[margin=1in]{geometry}
\usepackage[english]{babel}
\usepackage{amsmath}
\usepackage{wasysym}
\usepackage{amsthm, amssymb}
\usepackage{circledsteps}
\usepackage{pxfonts}
\usepackage{siunitx}
\usepackage{lmodern}

\newtheorem{hypothesis}{Hypothesis}
\newtheorem{result}{Result}
\usepackage{graphicx}
\usepackage{float}
\usepackage{natbib}
\bibpunct{(}{)}{;}{a}{,}{,}
\usepackage{booktabs}
\usepackage{comment} 
\usepackage{appendix}
\usepackage[flushleft]{threeparttablex}
\usepackage{array} % For fixed column width
\usepackage{ragged2e}
\usepackage{caption}
\usepackage{subfig}
\usepackage{setspace}
\usepackage{multirow}
\usepackage{color}
\usepackage{enumitem}
\usepackage{bbm}
\theoremstyle{definition}

\theoremstyle{plain}

\providecommand{\keywords}[1]
{\small\textbf{Keywords:} #1
}

\usepackage{hyperref}

\DeclareFontShape{TU}{lmr}{m}{scit}{<->ssub*lmr/m/sc}{}
\DeclareFontShape{TU}{lmr}{bx}{sc}{<->ssub*lmr/m/sc}{}
\DeclareFontShape{TU}{lmr}{bx}{scit}{<->ssub*lmr/m/sc}{}
\DeclareFontShape{TU}{lmr}{b}{sc}{<->ssub*lmr/m/sc}{}
\DeclareFontShape{TU}{lmr}{b}{scit}{<->ssub*lmr/m/sc}{}

\newcommand\Payment{\textit{\scshape{2}}}
\newcommand\ED{\textit{\scshape{ED}}}
\newcommand\D{\textit{\scshape{D}}}
\newcommand\EP{\textit{\scshape{EP}}}

\renewcommand{\P}{\textit{\scshape{P}}}

\title{
When Medical AI Explanations Help and When They Harm\footnotemark[1]}
\author{Manshu Khanna$^{a}$, Ziyi Wang$^{b}$, Lijia Wei$^{b}$$^{*}$, Lian Xue$^{b}$$^{*}$ \\
        \small $^{a}$HSBC Business School, Peking University, Shenzhen, China \\
        \small $^{b}$Economics and Management School, Wuhan University, Wuhan, China
}

\date{}

\renewcommand{\thefootnote}{\arabic{footnote}}

\begin{document}

\maketitle
\begin{abstract} 
\noindent 
We document a fundamental paradox in AI transparency: explanations improve decisions when algorithms are correct but systematically worsen them when algorithms err. In an experiment with 257 medical students making 3,855 diagnostic decisions, we find explanations increase accuracy by 6.3 percentage points when AI is correct (73\% of cases) but decrease it by 4.9 points when incorrect (27\% of cases). This asymmetry arises because modern AI systems generate equally persuasive explanations \emph{regardless} of recommendation quality—physicians cannot distinguish helpful from misleading guidance. We show physicians treat explained AI as 15.2 percentage points more accurate than reality, with over-reliance persisting even for erroneous recommendations. Competent physicians with appropriate uncertainty suffer most from the AI transparency paradox (-12.4pp when AI errs), while overconfident novices benefit most (+9.9pp net). Welfare analysis reveals that selective transparency generates \$2.59 billion in annual healthcare value, 43\% more than the \$1.82  billion from mandated universal transparency.
\end{abstract} 

\noindent \keywords{Artificial Intelligence (AI); Lab-in-the-field experiment; Belief updating; Bayesian. }

\renewcommand{\thefootnote}{*}
\footnotetext[1]{
We thank Qiuhao Chen, Jingru Cui, Pan Li, Xinyu Liang, and Yuqi Mou for their assistance with the experimental sessions. We also thank Bozhang Xia and Jin Xi for helpful proofreading and comments, and Lyu Chen for valuable discussions. Lijia Wei acknowledges support from the National Natural Science Foundation of China (Grant Nos.~72433003 and 72173093) and the Center for Behavioral and Experimental Research (CBER) at Wuhan University. Lian Xue acknowledges support from the National Natural Science Foundation of China (Grant No.~71903147). Manshu Khanna acknowledges financial support from the Peking University Digital and Humanities Special Grant. The experiment was preregistered on the AEA registry, number AEARCTR-0015153. All authors contributed equally and declare no conflict of interest. $^{*}$Co-corresponding authors emails: \tt{ljwei@whu.edu.cn; lianxue@whu.edu.cn}
}

\newpage

\section{Introduction}
\renewcommand{\thefootnote}{\arabic{footnote}}

Artificial intelligence increasingly determines consequential economic outcomes. In healthcare alone, AI systems influence 500 million annual diagnostic decisions globally, with the market projected to reach \$188 billion by 2030 \citep{GrandView2023}. This rapid deployment has prompted a regulatory revolution: the European Union's AI Act mandates explanations for high-risk systems, the FDA requires interpretability for medical AI, and China's 2024 guidelines standardize transparency requirements.\footnote{The EU AI Act (Article 13) \citep{EUAIAct2024} requires high-risk AI systems to be ``sufficiently transparent to enable deployers to interpret the system's output and use it appropriately" (Last accessed on 27 November 2025 at \url{https://artificialintelligenceact.eu/article/13/}). The FDA's Software as Medical Device framework mandates ``clinically relevant description of the device's logic.'' China's National Health Commission released comprehensive transparency guidelines in January 2024.} These regulations converge on a shared premise: understanding why an algorithm makes specific recommendations improves human decision-making.

We document the opposite. Using incentivized diagnostic decisions by medical students (prospect physicians), we establish what we term the \textit{AI Transparency Paradox}: explanations for algorithmic recommendations systematically worsen outcomes when algorithms err, creating near-symmetric effects in opposite directions that fundamentally challenge current regulatory approaches. When AI advice is correct explanations increase diagnostic accuracy, but when AI advice is incorrect explanations decrease accuracy. This swing reveals that explanations amplify algorithmic influence regardless of recommendation quality. More troublingly, explanations degrade decisions in over one-quarter of cases, precisely when human judgment matters most.\footnote{Given realistic AI accuracy levels of 70-80\% in medical applications \citep{nori2023capabilities, singhal2023large}, this creates a fundamental policy tension that current regulations fail to address.}

The paradox stems from a fundamental feature of modern AI. Language models like GPT-4 generate explanations by pattern-matching medical knowledge to their observations, producing equally fluent reasoning whether diagnosing correctly or incorrectly.\footnote{A technical feature fundamentally different from human communication, where explanation quality naturally signals reliability through linguistic cues like disfluencies and hedging \citep{graeber2024explanations}.} The explanation ``ST-segment elevation indicates myocardial ischemia, consistent with acute coronary syndrome'' sounds authoritative regardless of whether the patient actually has a heart attack or pericarditis mimicking these symptoms. Humans, however, treat explanation quality as a signal of accuracy. A compelling medical reasoning suggests a correct diagnosis. They apply what we formalize as a persuasiveness weight $\psi(e) = \exp(\lambda \cdot q(e))$ that inflates their posterior beliefs proportional to explanation quality. Our data reveal physicians update as if explained AI has 88.2\% accuracy when correct (15.2 percentage points above the true 73\%) and 79.2\% accuracy when incorrect (still 6.2 points above truth). This systematic miscalibration, which we measure through both anticipated trust before seeing advice and revealed accuracy from actual belief updating, drives the paradox.

To establish this paradox, we conduct a lab-in-the-field experiment with 257 medical students at a major Chinese teaching hospital, generating 3,855 incentivized diagnostic decisions across 15 clinical scenarios. Our $2 \times 2$ factorial design crosses two dimensions: whether AI advice includes explanatory reasoning (with versus without explanation) and how uncertainty is communicated (deterministic versus probabilistic format). Participants complete three sequential tasks for each scenario: (1) providing initial diagnostic assessments before seeing AI advice, (2) predicting how AI advice will affect their confidence, and (3) updating their assessments after receiving AI recommendations generated by GPT-4o with 73.4\% accuracy. We use quadratic scoring rules to ensure truthful belief reporting and measure both the level of diagnostic accuracy and the dynamics of belief updating. This three-stage elicitation enables us to identify both ex-ante trust (anticipated learning from AI) and ex-post implied accuracy (revealed trust from actual belief updating), providing theory-grounded measures of over-reliance that distinguish our approach from prior work.

Our experimental results establish five key findings. First, we document the core paradox: when AI advice is correct (73\% of cases), explanations increase diagnostic accuracy by 6.3 percentage points ($p<0.01$). When AI advice is incorrect (27\% of cases), explanations decrease accuracy by 4.9 percentage points ($p<0.01$). The near-symmetric magnitudes reveal that explanations amplify AI influence regardless of recommendation quality, creating an 11.2 percentage point paradox. Given realistic AI accuracy, the net welfare effect is positive but modest (+3.3 percentage points)---far smaller than examining correct cases alone would suggest, achieving only 52\% of the theoretical first-best outcome.

Second, the problem intensifies rather than resolves with conventional mitigation strategies. Prior research suggests communicating uncertainty through probabilistic outputs should protect against over-reliance \citep{kompa2021second, zhang2020effect}. We test this by crossing explanation presence with advice format: deterministic (``AI recommends B'') versus probabilistic (``70\% B, 30\% A''). Surprisingly, probabilistic formats amplify the harm: when AI errs, explanations decrease accuracy by 6.6 percentage points with probabilistic presentation versus only 3.1 percentage points ($p<0.05$) with deterministic. Rather than signaling caution, probabilistic outputs create ambiguity that explanations then resolve into a \emph{false certainty}. A physician seeing ``70\% likelihood of pneumonia'' might appropriately maintain skepticism, but add a convincing explanation of consolidation patterns and bacterial markers, and that expressed uncertainty transforms into misplaced confidence.

Third, we identify the core mechanism driving these effects: the symmetric persuasiveness of AI-generated explanations creates asymmetric welfare outcomes. Modern language models produce explanations conditional only on their observations, not ground truth: $\Pr(e|x,\theta) = \Pr(e|x)$. This means explanation quality is identical whether the AI is correct or incorrect: $\mathbb{E}[q(e)|x=\theta] = \mathbb{E}[q(e)|x\neq\theta]$. However, physicians treat explanation quality as diagnostic of recommendation accuracy, applying a persuasiveness weight $\psi(e,x,\theta) = \exp(\lambda \cdot q(e))$ that inflates their posterior beliefs. This mismatch creates systematic over-reliance: physicians update as if explained AI has 88.2\% accuracy when correct (15.2 percentage points above the true 73\%) and 79.2\% accuracy when incorrect (6.2 points above truth). Critically, this over-reliance persists even for erroneous recommendations, operating through both ex-ante expectations (anticipated learning increases by 3.3 pp) and ex-post behavior (revealed trust increases by 4.3 pp). This differs fundamentally from human-to-human communication studied by \citet{graeber2024explanations}, where explanation quality contains natural signals of reliability that help receivers assess accuracy.

Fourth, this symmetric persuasiveness operates through three complementary mechanisms at different cognitive layers. \textit{Over-reliance} (calibration failure) inflates perceived AI reliability regardless of actual accuracy, with 83.7\% of physicians exhibiting over-reliance when AI is correct and 76.6\% when incorrect. \textit{Discernment failure} (signal processing) impairs physicians' ability to separate good advice from bad: explanations increase belief revision by +0.084 for incorrect advice versus only +0.030 for correct advice, eroding differential responses to signal quality. \textit{False confidence} (metacognitive failure) inflates diagnostic certainty by 4.6 percentage points even when AI is wrong, creating dangerous overconfidence precisely when caution is most needed. Explanations increase confidence by 4.6 percentage points even when AI is incorrect (from 0.840 to 0.886, $p<0.01$), creating false certainty precisely when caution is most needed.

Fifth, the paradox creates perverse distributional effects through confidence-competence misalignment. Explanations provide the greatest benefits (+9.9 pp) to overconfident novices---low-competence physicians who are wrong but certain---by leveraging their receptivity to authoritative guidance for corrective purposes. Conversely, explanations offer minimal gains (+3.6 pp) to humble experts---high-competence physicians with appropriate uncertainty---while causing substantial harm when AI errs (-12.4 pp). The benefit-to-harm ratio differs by 14-fold: 4.03 for overconfident novices versus 0.29 for humble experts. This pattern rewards unjustified confidence over sound clinical judgment, amplifying rather than mitigating skill gaps.

These findings carry immediate policy implications. Using conservative estimates of 500 million annual AI-assisted diagnoses globally and \$11,000 per diagnostic error \citep{tehrani2013diagnostic}, we evaluate alternative transparency policies through counterfactual welfare analysis. Universal transparency mandates---the approach embodied in the EU AI Act---generate \$1.82 billion in annual value but achieve only 52\% efficiency relative to the first-best benchmark. In contrast, contingent transparency policies that selectively provide explanations achieve dramatically higher welfare. Confidence-based policies (explanations only when AI confidence exceeds the 85th percentile) reach 70\% efficiency (\$2.42B value), while competence-adaptive policies (targeting explanations to low-competence physicians who benefit from correction while protecting high-competence physicians from misleading advice) achieve 75\% efficiency (\$2.59B value)---43\% higher than universal transparency. These gains can extend to other high-stakes domains where algorithmic errors carry substantial costs: criminal sentencing \citep{kleinberg2018human}, lending decisions \citep{fuster2022predictably}, and hiring \citep{cowgill2020biased}.

The deeper implication challenges the fundamental assumption underlying current regulatory approaches: that more information necessarily improves decisions. Our findings reveal that \textit{strategic silence}---withholding explanations when they would mislead---represents optimal AI assistance. However, modern AI systems cannot implement this wisdom because they maintain uniform explanation fluency regardless of underlying reliability. As language models advance (GPT-5, GPT-6), this challenge intensifies: better explanation generation amplifies the paradox rather than resolving it. The decoupling of explanation quality from recommendation accuracy means that even as AI systems become more capable of generating sophisticated, literature-citing, professorial explanations, these explanations become more persuasive for both correct and incorrect recommendations alike.

\bigskip

\paragraph{Related Literature. } Our findings make three primary contributions to economics and the broader literature on human-AI collaboration. First, we provide the first causal evidence that transparency can systematically reduce welfare when information sources are fallible. While information design theory establishes conditions under which senders benefit from disclosure \citep{kamenica2011bayesian, blackwell1953equivalent}, we show that mechanically generated explanations---lacking strategic optimization---can harm receivers through symmetric amplification of both correct and incorrect signals. This extends information economics to settings where signal richness and signal value can diverge: explanatory content from fallible sources reduces welfare by inflating perceived signal quality regardless of actual reliability.

Second, we identify a novel mechanism---explanation-induced over-reliance operating through symmetric persuasiveness---that differs fundamentally from existing accounts of algorithm aversion \citep{dietvorst2015algorithm} and appreciation \citep{logg2019algorithm}. Our theory-grounded measures ($\alpha$ for ex-ante trust and $\mu^{\text{implied}}$ for ex-post revealed accuracy) reveal that explanations increase perceived AI reliability by 15 percentage points through dual channels: inflated expectations before seeing advice and miscalibrated beliefs after updating. This bias persists even for incorrect recommendations because modern AI systems generate explanations conditional only on observations, not ground truth---creating uniform persuasiveness that humans cannot appropriately discount. This mechanism contrasts sharply with human communication studied by \citet{graeber2024explanations}, where explanation quality naturally contains reliability signals through linguistic markers, enabling receivers to assess accuracy.

Third, we challenge the regulatory consensus on algorithmic transparency embodied in the EU AI Act, U.S. FDA guidelines, and emerging Chinese frameworks. Current regulations mandate explanations for high-risk AI systems under the assumption that understanding improves collaboration. Our evidence suggests this one-size-fits-all approach imposes substantial welfare costs---\$1.6 billion annually in healthcare alone. We demonstrate that contingent policies targeting explanations based on AI confidence or user competence achieve near-optimal welfare while remaining administratively feasible. These findings suggest three concrete regulatory reforms: (1) permitting conditional transparency based on algorithmic confidence thresholds, (2) enabling competence-adaptive systems that adjust explanation provision based on user expertise, and (3) requiring impact evaluation through randomized trials rather than mandating specific transparency mechanisms.

Our results also contribute to the adjacent literature. For explainable AI research \citep{arrieta2020explainable, holzinger2017we}, we provide causal evidence that interpretability can worsen decisions when algorithms err---a possibility largely absent from computer science literature driving transparency requirements. For medical decision-making \citep{croskerry2003decision}, we show explanations create new anchors rather than correcting existing biases. For human-AI interaction \citep{buccinca2021trust}, we demonstrate that explanation effects depend critically on the interaction between AI quality, advice format, and user expertise, challenging simple prescriptions for transparency design. Our findings complement recent evidence on human-AI teaming reliability \citep{liu2025human}, which demonstrates that while AI augments clinician performance, human-machine teams rarely achieve full complementarity. This aligns with our transparency paradox: explanations improve outcomes conditionally rather than universally.

\bigskip
The paper proceeds as follows. Section~\ref{sec:design} details the experimental design, including the medical scenarios, AI advice generation, and belief elicitation procedures. Section~\ref{sec:framework} introduces the Bayesian framework to derive behavioral predictions and operationalize our theory-grounded measures of over-reliance. Section~\ref{sec:results} presents the main experimental results, establishing the transparency paradox and testing the underlying mechanisms. Section~\ref{sec:conclusion} concludes.

\section{Experimental Design and Procedure}
\label{sec:design}

\subsection{Experimental Design}

We conducted a lab-in-the-field experiment with 257 medical students to examine how AI-generated diagnostic advice affects belief updating. We employed a 2×2 between-subjects factorial design crossing two dimensions: (1) whether AI advice includes explanatory reasoning (with vs. without explanation) and (2) how AI advice is presented (deterministic vs. probabilistic format).

In all treatments, participants encountered 15 \emph{medical prescription scenarios} requiring them to prescribe the appropriate medication from five options based on patient symptoms and history. All scenarios were drawn from a validated test bank of 817 questions used in Chinese medical education and practitioner exams, with each scenario having a single correct treatment option.
Participants received AI advice in their assigned format during each scenario. In the deterministic format, AI provided a single recommendation (``AI recommends medication B''); in the probabilistic format, AI provided its top two choices with probabilities (``AI recommends: 70\% medication B, 30\% medication A'').

When explanations were provided, AI advice included medical reasoning. For example, in the Explained Probabilistic (\EP) condition:

\begin{quote}
\small
\textit{
Male, 68 years old, with a 5-year history of prior myocardial infarction. He has been taking perindopril, metoprolol, and aspirin regularly, with no episodes of chest pain. Physical examination is unremarkable. Laboratory tests show: total cholesterol (TC) 5.0 mmol/L, LDL-C 2.9 mmol/L, triglycerides (TG) 5.9 mmol/L, HDL-C 0.9 mmol/L. What is the preferred lipid-lowering medication for this patient? \\
A. Ezetimibe \\
B. Atorvastatin \\
C. Colestipol \\
D. Rosuvastatin \\
E. Fenofibrate\\
}

\textit{AI Advice: AI recommends 70\% medication B (Atorvastatin), 30\% medication E (Fenofibrate).}

\textit{AI Explanation: The patient is 68 years old with a history of old myocardial infarction, currently on standard medication therapy, and has no episodes of chest pain. Laboratory results show: LDL-C: 2.9 mmol/L (slightly above the recommended target <1.8 mmol/L); TG: 5.9 mmol/L (significantly elevated, recommended <1.7 mmol/L); HDL-C: 0.9 mmol/L (low, recommended $\geq$1.0 mmol/L). Considering the history of myocardial infarction and severe hypertriglyceridemia, the patient falls into the very high-risk category, requiring further LDL-C reduction and TG control. B.Atorvastatin(approx. 70\% likelihood of being correct) is a potent HMG-CoA reductase inhibitor that can significantly lower LDL-C and also has some effect in reducing triglycerides. It is one of the first-line drugs for treating coronary heart disease and very high-risk patients, with proven benefit in improving prognosis. E.Fenofibrate(approx. 30\% likelihood of being correct), a fibrate drug, is particularly effective for severe hypertriglyceridemia but has weaker effects on lowering LDL-C. For very high-risk patients whose LDL-C remains above target, statins should be prioritized. If TG levels remain unsatisfactory, fenofibrate may be considered in combination therapy.}
\end{quote}

\begin{figure}[h!]
    \centering  
    \includegraphics[width=\textwidth]{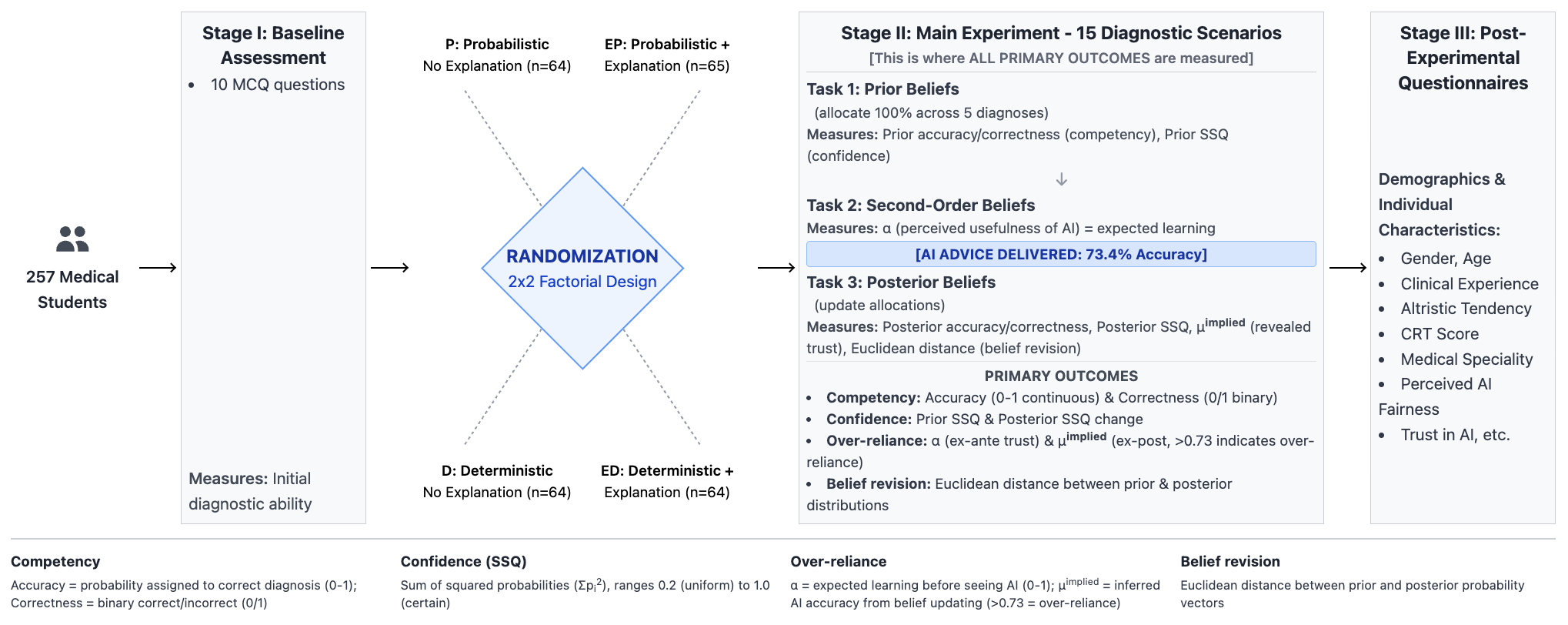}
    \caption{Experimental Design and Measurement Framework}
    \caption*{\footnotesize \textit{Notes:} 257 medical students were randomized into a 2×2 between-subjects factorial design. All participants completed identical Stage I (baseline) and Stage III (questionnaires). In Stage II, participants completed 15 clinical scenarios with three belief elicitation tasks each. AI advice was delivered between Tasks 2 and 3, with format varying by treatment condition.}
    \label{fig:experimental_design}
\end{figure}

This yields four treatment conditions: Probabilistic (\P), Explained Probabilistic (\EP), Deterministic (\D), and Explained Deterministic (\ED). Figure~\ref{fig:experimental_design} illustrates the experimental structure.

The experiment consisted of three stages:

\begin{itemize}
\item \textbf{Stage I: Ability assessment.} Participants completed 10 multiple-choice medical prescription questions, earning 1 yuan per correct answer.\footnote{These questions were drawn from the same test bank used for the main experiment's medical prescription scenarios in Stage II.} Their average accuracy was 61.7\%, establishing a baseline below the AI's 73.4\% performance. Following Stage I, participants were randomly assigned to the four treatment conditions (N=64-65 per condition).

\item \textbf{Stage II: Main experiment (15 medical prescription scenarios).} Stage II constituted our main experimental stage, where each participant encountered 15 medical prescription scenarios randomly drawn from the test bank. Each scenario involved three sequential tasks:
    \begin{itemize}
    \item \textit{Task 1 (prior belief elicitation):} Participants read the medical prescription scenario and allocated 100 percentage points across five possible treatment options based on their assessment. The sum of squared probabilities (SSQ) captures the concentration of their beliefs, ranging from 0.2 (complete uncertainty: 20\% on each option) to 1.0 (complete certainty: 100\% on one option). We presented SSQ to participants as an ``\emph{informativeness score}'', measuring how concentrated—and thus informative—their probability distribution was about the correct option.
    
    \item \textit{Task 2 (second-order belief elicitation):} Before seeing AI advice, participants predicted how their informativeness score (SSQ) would change after receiving AI advice. For example, if they expected to move from uniform beliefs (20\% each, SSQ=0.2) to concentrated beliefs (80\% on one option, SSQ=0.68), they would predict a substantial increase, indicating high anticipated learning from the AI.
    
    \item \textit{Task 3 (posterior belief elicitation):} After receiving AI advice with 73.4\% accuracy, formatted according to their treatment condition, participants updated their probability allocations. They completed the same probability assignment interface as in Task 1, now incorporating the AI advice into their assessment.
    \end{itemize}
    
\item \textbf{Stage III: Post-Experimental Questionnaires.} In Stage III, we collected demographic information (gender, age, clinical experience), cognitive ability measures (CRT \citep{frederick2005cognitive}), altruistic tendencies, medical specialty, and AI-related attitudes through structured questionnaires (detailed in Appendix~\ref{sec:append_instructions}).
\end{itemize}

\subsection{Implementation}

We recruited 257 medical students in clinical years (4th-6th) from Zhongnan Hospital's affiliated medical school.%
\footnote{Zhongnan Hospital (Central South Hospital) is one of the largest hospitals in Hubei Province, recognized for its contributions to medical research and diverse patient population. Recruitment was facilitated by the hospital research deputy through general invitations and advertisements. Participants were informed the study examined medical decision-making and was conducted by Wuhan University's Centre for Behavioural and Experimental Research (CBER).} Sessions accommodated approximately 20 participants each in conference rooms, lasting 60 minutes, using oTree \citep{chen2016otree}. Treatment assignment occurred through random card draws; Table~\ref{tab:balancecheck} confirms successful randomization across all observable characteristics (F-test $p>0.10$). The experiment was preregistered at the AEA RCT Registry in January 2025 \citep{belief2025}.

The 15 medical prescription scenarios were randomly selected from a validated test bank of 817 cases used in Chinese medical education and practitioner examinations. ChatGPT-4o generated AI advice for all 817 scenarios, achieving 73.4\% accuracy, which is significantly higher than participants' 61.7\% baseline accuracy in Stage I, yet remaining realistically imperfect.\footnote{This accuracy level aligns with recent medical AI benchmarks \citep{nori2023capabilities, singhal2023large}, providing sufficient variation to study both beneficial and harmful effects of AI explanations.} This performance gap ensures participants could potentially benefit from AI advice while the AI's fallibility allows us to examine the transparency paradox when recommendations are incorrect.

Payment followed incentive-compatible quadratic scoring rules to ensure truthful belief reporting \citep{hossain2013binarized}. For probability distributions $\mathbf{p} = (p_1, ..., p_5)$ in Tasks 1 and 3 (Stage II), participants earned $\Pi(\mathbf{p}) = 2 - \sum_j (p_j - x_j)^2$ yuan, where $x_j = 1$ for the correct option and 0 otherwise. Task 2 rewarded accurate SSQ predictions with up to 2 yuan. Each scenario thus offered maximum earnings of 6 yuan (2 yuan × 3 tasks), totaling 90 yuan across all scenarios. Combined with Stage I earnings (up to 10 yuan), Stage III cognitive tasks and the participation payment (30 yuan), the average total payment reached 108 yuan (\$15 USD), which is approximately five times the local hourly minimum wage.%
\footnote{At the experimental exchange rate of 7.2 yuan/USD, payments were substantial relative to the local minimum wage of 22 yuan/hour \citep{WuhanMinWage2020}.} 

Beyond individual incentives, we implemented a patient donation mechanism to simulate real medical stakes where physician decisions affect patient welfare \citep{godager2013profit,currie2017diagnosing}. Higher diagnostic accuracy generated larger charitable contributions to ALS patients (up to 10 yuan per scenario), with total donations of 1,929.87 yuan (\$268 USD) made on behalf of participants. The donation certificate is provided in Appendix Figure~\ref{fig:donation}.

\subsection{Primary Outcome Measures}

Our experiment elicits probability distributions $\mathbf{p}_i = (p_{i,1}, ..., p_{i,5})$ over five diagnostic options at two time points: before (Task 1) and after (Task 3) receiving AI advice. From these elicited beliefs, we construct measures capturing both the level and dynamics of diagnostic assessments.\footnote{All primary outcome measures derive from Stage II data. Stage I provides baseline ability controls, while Stage III supplies individual characteristic controls used in heterogeneity analyses.}

\textit{Confidence and competence measures.} We measure physicians' subjective certainty through the sum of squared probabilities (SSQ):
\begin{equation*}
\text{SSQ}_i = \sum_{j=1}^{5} p_{i,j}^2
\end{equation*}
ranging from 0.2 (complete uncertainty: equal 20\% on each option) to 1.0 (complete certainty: 100\% on one option). Diagnostic quality is captured through two complementary metrics. \textit{Accuracy} ($A$) measures the probability assigned to the correct diagnosis: $A_i = p_{i,\text{correct}}$, providing a continuous measure of how much weight physicians place on the true option. \textit{Correctness} ($C$) is a binary indicator of whether the physician's modal (highest probability) choice matches the true diagnosis: $C_i = \mathbbm{1}[\arg\max_j p_{i,j} = \text{correct}]$, reflecting the final diagnostic decision.

\textit{Belief dynamics.} We quantify the magnitude of belief updating through belief revision, calculated as the Euclidean distance between prior and posterior probability vectors:
\begin{equation*}
d(\mathbf{p}_{\text{prior}}, \mathbf{p}_{\text{post}}) = \sqrt{\sum_{j=1}^{5}(p_{j,\text{post}} - p_{j,\text{prior}})^2}
\end{equation*}
This metric, ranging from 0 to $\sqrt{2}$, captures the overall magnitude of the belief shift induced by AI advice.

These measures are incentivized through quadratic scoring rules, ensuring truthful belief reporting (see Section 2.2 for payment details). The theoretical interpretation and additional theory-grounded measures derived from these elicited beliefs are developed in Section 3.3.

% ============================================================
% SECTION 3: Conceptual Framework and Behavioral Hypotheses
% ============================================================

\section{Conceptual Framework and Behavioral Hypotheses}
\label{sec:framework}

\subsection{Bayesian Framework}

Consider an artificial intelligence (AI) system that observes a noisy signal about an unknown medical condition and must communicate this information to a physician decision maker. Let $\theta \in \Theta = \{\theta_1, \theta_2, ..., \theta_n\}$ denote the true diagnosis, where $n = 5$ in our experimental setting. The physician holds a prior belief $p \in \Delta(\Theta)$ over possible diagnoses based on their initial assessment.

The AI observes a private signal $x \in X = \Theta$ about the true diagnosis with known accuracy $\mu = 0.73$. The signal structure follows:
\begin{equation*}
\Pr(x = \theta' | \theta) = \begin{cases}
\mu & \text{if } \theta' = \theta \\
\frac{1-\mu}{n-1} & \text{if } \theta' \neq \theta
\end{cases}
\end{equation*}
This represents a symmetric error structure where the AI correctly identifies the true diagnosis with probability $\mu$ and, when wrong, is equally likely to suggest any other diagnosis.

The AI communicates with the physician using one of four protocols: (1) Deterministic (\D{}), sending a single recommendation $m^D = x \in \Theta$; (2) Probabilistic (\P{}), sending its posterior distribution $m^P = q \in \Delta(\Theta)$ where $q_j = \mu$ if $j = x$ and $q_j = \frac{1-\mu}{n-1}$ otherwise; (3) Deterministic with Explanation (\ED{}), sending $(m^D, e)$ where $e \in E$ is a natural language explanation; and (4) Probabilistic with Explanation (\EP{}), sending $(m^P, e)$.

A crucial observation is that protocols \D{} and \P{} convey identical information about the AI's observation. Since the AI's accuracy $\mu$ is known and fixed, the mapping from observation $x$ to posterior distribution $q$ is one-to-one. When the physician receives either $m^D = \theta_i$ or $m^P$ with $q_i = \mu$, they can perfectly infer that the AI observed $x = \theta_i$.

Given the AI's signal, the optimal Bayesian posterior is identical under both deterministic and probabilistic protocols:
\begin{equation}
\pi^{\text{Bayes}}(\theta_j | x = \theta_i) = \begin{cases}
\frac{p(\theta_i) \cdot \mu}{p(\theta_i) \cdot \mu + \sum_{k \neq i} p(\theta_k) \cdot \frac{1-\mu}{n-1}} & \text{if } j = i \\
\frac{p(\theta_j) \cdot \frac{1-\mu}{n-1}}{p(\theta_i) \cdot \mu + \sum_{k \neq i} p(\theta_k) \cdot \frac{1-\mu}{n-1}} & \text{if } j \neq i
\end{cases}
\label{eq:bayes_posterior}
\end{equation}
This normative benchmark assumes the physician correctly processes the AI's known accuracy and updates beliefs using Bayes' rule. Any systematic deviation from this posterior constitutes a behavioral bias.

\subsection{Explanation-Augmented Belief Updating}

While protocols \D{} and \P{} are informationally equivalent, adding explanations fundamentally alters belief updating. We now develop a model explaining why explanations create a paradox: improving accuracy when AI advice is correct but harming it when incorrect.

When physicians receive explanations, their posterior becomes:
\begin{equation}
\pi^E(\theta | x, e) = \frac{p(\theta) \cdot L(x, e | \theta)}{\sum_{\theta'} p(\theta') \cdot L(x, e | \theta')}
\label{eq:posterior_explained}
\end{equation}
where $L(x, e | \theta)$ is the perceived likelihood function. The key departure from optimal updating is that physicians cannot perfectly assess explanation quality. We decompose this as:
\begin{equation*}
L(x, e | \theta) = \Pr(x | \theta) \cdot \psi(e, x, \theta)
\end{equation*}
where $\psi(e, x, \theta)$ represents the persuasiveness weight from the explanation.

The paradox arises from a fundamental mismatch. AI systems generate explanations conditional only on their observation, not on ground truth:
\begin{equation*}
\Pr(e | x, \theta) = \Pr(e | x)
\end{equation*}
This means explanations are equally fluent and convincing whether the AI is correct ($x = \theta$) or incorrect ($x \neq \theta$). Modern language models can generate plausible-sounding reasoning for any diagnosis they recommend.

However, physicians perceive explanation quality as informative about accuracy. They apply a persuasiveness function:
\begin{equation}
\psi(e, x, \theta) = \exp(\lambda \cdot q(e))
\label{eq:persuasiveness}
\end{equation}
where $q(e) \in [0, 1]$ measures the perceived quality of explanation $e$, and $\lambda > 0$ captures the physician's susceptibility to explanations.

The critical insight is that while physicians treat $q(e)$ as diagnostic of whether $x = \theta$, in reality:
\begin{equation*}
\mathbb{E}[q(e) | x = \theta] = \mathbb{E}[q(e) | x \neq \theta] = \bar{q}
\end{equation*}
Expected explanation quality is identical for correct and incorrect recommendations.

This mechanism creates systematic over-weighting. The physician's posterior for the AI-recommended diagnosis becomes:
\begin{equation}
\pi^E(\theta_i | x = \theta_i, e) = \frac{p(\theta_i) \cdot \mu \cdot \exp(\lambda \cdot q(e))}{p(\theta_i) \cdot \mu \cdot \exp(\lambda \cdot q(e)) + \sum_{k \neq i} p(\theta_k) \cdot \frac{1-\mu}{n-1}}
\label{eq:posterior_overweight}
\end{equation}

The explanation increases the weight on the AI's recommendation by factor $\exp(\lambda \cdot q(e))$, regardless of correctness. This creates asymmetric effects on diagnostic accuracy. Define accuracy as $\Pr(a = \theta)$, where the physician chooses action $a$ corresponding to their highest posterior belief:
\begin{equation}
\text{Accuracy} = \mu \cdot \Pr(a = \theta | x = \theta) + (1-\mu) \cdot \Pr(a = \theta | x \neq \theta)
\label{eq:accuracy_decomp}
\end{equation}

With explanations, when the AI is correct ($x = \theta$), over-weighting increases $\pi^E(\theta | x = \theta)$, making the physician more likely to select the correct diagnosis. When the AI is incorrect ($x \neq \theta$), the same over-weighting increases $\pi^E(x | x \neq \theta)$ for the wrong diagnosis, making the physician less likely to select the true diagnosis $\theta$. The explanation effectively makes physicians treat the AI as having higher accuracy than $\mu = 0.73$, regardless of whether the specific recommendation is correct.

\subsection{Operationalizing Over-Reliance: Theory-Grounded Measures}

Our theoretical framework predicts that explanations induce systematic over-reliance on AI advice by inflating physicians' perception of its accuracy. To test this prediction empirically, we construct two theory-grounded measures that reveal physicians' actual versus anticipated trust in AI recommendations.

\subsubsection*{Ex-ante Trust: Anticipated Learning from AI ($\alpha$)}

Before physicians see AI advice in Task 2, we elicit their \textit{second-order beliefs}---their expectations about how their confidence will change after receiving the recommendation. This allows us to measure their ex-ante trust: how much they anticipate learning from the AI before observing its actual advice.

Following the restricted belief updating framework of \citet{chambers2021dynamic}, we model physicians' anticipated learning as a mixture: with probability $\alpha$, they expect the AI will be perfectly informative (revealing the true diagnosis with certainty, yielding $\text{SSQ} = 1$); with probability $(1-\alpha)$, they expect to learn nothing new (SSQ remains at the prior level). This yields:
\begin{equation*}
\mathbb{E}[\text{SSQ}_{\text{posterior}}] = \alpha \cdot 1 + (1-\alpha) \cdot \text{SSQ}_{\text{prior}}
\end{equation*}

Solving for the trust parameter:
\begin{equation}
\alpha = \frac{\text{SSQ}_{\text{expected}} - \text{SSQ}_{\text{prior}}}{1 - \text{SSQ}_{\text{prior}}}
\label{eq:alpha}
\end{equation}
where $\text{SSQ}_{\text{prior}} = \sum_j p_{j,\text{prior}}^2$ is the physician's initial confidence and $\text{SSQ}_{\text{expected}}$ is their stated expectation (elicited in Task 2) of posterior confidence. The parameter $\alpha \in [0,1]$ captures the ex-ante probability that physicians believe the AI will provide fully informative advice.

\subsubsection*{Ex-post Implied Accuracy: Revealed Trust from Behavior ($\mu^{\text{implied}}$)}

While $\alpha$ captures expectations, physicians' actual belief updating reveals their implicit assessment of AI accuracy. From the Bayesian posterior in equation \eqref{eq:bayes_posterior}, we can infer what AI accuracy $\mu^{\text{implied}}$ would rationalize the observed belief shifts if physicians were perfect Bayesians.

Given a physician's prior $p_k$ on option $k$ and posterior $\pi(\theta_k)$ after AI recommends $k$, we reverse-engineer:
\begin{equation}
\mu^{\text{implied}} = \frac{\pi(\theta_k) \cdot \sum_{j \neq k} p(\theta_j) / (n-1)}{p(\theta_k)(1-\pi(\theta_k)) + \pi(\theta_k) \cdot \sum_{j \neq k} p(\theta_j) / (n-1)}
\label{eq:mu_implied}
\end{equation}

This formula answers: ``If this physician were a perfect Bayesian updater, what AI accuracy would she need to believe in to produce her observed posterior?'' When $\mu^{\text{implied}} > 0.73$ (the true AI accuracy), the physician exhibits over-reliance---she updates as if the AI were more accurate than it actually is.

\textit{Intuition through example.} Suppose a physician starts with $p_B = 0.30$ and updates to $\pi_B = 0.85$ after AI recommends B. This large shift implies $\mu^{\text{implied}} \approx 0.85$ (85\%), far exceeding the true 73\%. She is treating the AI recommendation as if it were 12 percentage points more reliable than reality.

\subsubsection*{Connecting Measures to Mechanisms}

These two measures capture complementary facets of the transparency paradox. Ex-ante trust ($\alpha$) tests whether explanations prime physicians to expect greater learning, reflecting the anticipatory channel through which explanations shape receptivity. Ex-post implied accuracy ($\mu^{\text{implied}}$) tests whether explanations actually inflate the weight physicians place on AI recommendations during belief updating, operationalizing the parameter $\lambda$ in equation \eqref{eq:persuasiveness}.

The divergence between anticipated ($\alpha$) and revealed ($\mu^{\text{implied}}$) trust is itself informative. If physicians systematically underestimate $\alpha$ relative to $\mu^{\text{implied}}$, they fail to anticipate how persuasive explanations will prove in practice---suggesting limited metacognitive awareness of explanation effects.

Table~\ref{tab:hypothesis_mapping} summarizes how these measures map to our behavioral hypotheses, which will be detailed in the next section.

\begin{table}[H]
\centering
\caption{Mapping Theory to Empirical Tests}
\label{tab:hypothesis_mapping}
\begin{threeparttable}
\begin{tabular}{lll}
\toprule
Hypothesis & Empirical Test & Measure \\
\midrule
H1: Paradox & $\Delta A|_{x = \theta} > 0$, $\Delta A|_{x \neq \theta} < 0$ & Accuracy ($A$) \\
H2: Probabilistic mitigation & $|\Delta A^{\text{Prob}}|_{x \neq \theta} < |\Delta A^{\text{Det}}|_{x \neq \theta}$ & Accuracy by format \\
H3a: Ex-ante trust & $\alpha^E > \alpha^{NE}$ & $\alpha$ \\
H3b: Ex-post trust & $\mu^{\text{impl},E} > \mu^{\text{impl},NE}$; persists when $x \neq \theta$ & $\mu^{\text{implied}}$ \\
H4: discernment & $|\text{Rev}|_{x=\theta}^E - |\text{Rev}|_{x \neq \theta}^E < |\text{Rev}|_{x=\theta}^{NE} - |\text{Rev}|_{x \neq \theta}^{NE}$ & Belief Revision \\
H5: False confidence & $\text{SSQ}_{\text{post}}^E > \text{SSQ}_{\text{post}}^{NE}$ when $x \neq \theta$ & SSQ \\
\bottomrule
\end{tabular}
\begin{tablenotes}
\small
\item \textit{Notes:} $E$ = Explanation condition; $NE$ = No Explanation; Prob = Probabilistic format; Det = Deterministic format. $A$ = Accuracy; Rev = Revision magnitude; SSQ = Sum of squared probabilities (confidence). H2 tests whether uncertainty communication mitigates explanation harm. H3b's persistence test examines whether over-reliance remains when AI is incorrect, the critical prediction distinguishing our symmetric persuasiveness mechanism from rational updating models.
\end{tablenotes}
\end{threeparttable}
\end{table}

\subsection{Behavioral Hypotheses}

We now formalize five testable predictions that follow from our theoretical framework.

\begin{hypothesis}[The AI Transparency Paradox]
\label{hyp:paradox}
Explanations create asymmetric effects on diagnostic accuracy: they improve accuracy when AI advice is correct ($x = \theta$) but systematically harm it when AI advice is incorrect ($x \neq \theta$).
\end{hypothesis}

This follows directly from equation \eqref{eq:accuracy_decomp}: explanations add persuasiveness weight $\exp(\lambda \cdot q(e))$ that does not discriminate between correct and incorrect advice. The \emph{symmetric} persuasiveness of AI-generated explanations, where $\mathbb{E}[q(e)|x=\theta] = \mathbb{E}[q(e)|x\neq\theta]$, creates \emph{asymmetric} welfare effects by amplifying both good and bad advice indiscriminately.

Next, we test the conventional hypothesis that quantifying uncertainty in AI recommendations can protect users from over-reliance, since being explicitly informed about the AI’s confidence helps them recognize when its advice may be unreliable \citep{kompa2021second,zhang2020effect}. Consequently, uncertainty communication may mitigate the harmful impact of explanations when AI advice is incorrect. Specifically, we examine whether $|\Delta A^{\text{Prob}}|_{x \neq \theta}| < |\Delta A^{\text{Det}}|_{x \neq \theta}|$, where $\Delta A$ denotes the explanation's effect on accuracy in probabilistic versus deterministic formats.

\begin{hypothesis}[Probabilistic Format as Mitigation]
\label{hyp:probabilistic}
Providing explanations with a probabilistic AI advice format (expressing uncertainty through probability distributions) mitigates the harmful effects of explanations when AI advice is incorrect. By explicitly signaling uncertainty, probabilistic outputs should help physicians appropriately discount potentially erroneous advice.
\end{hypothesis}

Having established the core paradox and tested a natural mitigation strategy, we now examine three mechanisms through which explanations operate.

\begin{hypothesis}[Over-reliance Mechanism]
\label{hyp:overreliance}
Explanations induce systematic over-reliance on AI advice through two channels: (a) ex-ante trust increases ($\alpha^E > \alpha^{NE}$), and (b) ex-post implied accuracy exceeds true accuracy ($\mu^{\text{implied}, E} > 0.73$). Critically, (c) this over-reliance persists even when AI advice is incorrect ($x \neq \theta$).
\end{hypothesis}

This hypothesis tests whether the persuasiveness function in equation \eqref{eq:persuasiveness} with $\lambda > 0$ empirically holds, and whether it applies symmetrically to correct and incorrect recommendations. The persistence of over-reliance when AI is incorrect, i.e., the third prediction, is critical: it distinguishes our symmetric persuasiveness mechanism from alternative accounts where users rationally update based on realized advice quality. Because modern language models generate explanations conditional only on observations ($\Pr(e|x)$, not $\Pr(e|x,\theta)$), physicians cannot identify when explanations are misleading, leading to persistent miscalibration.

\begin{hypothesis}[Impaired Discernment]
\label{hyp:discrimination}
Explanations impair physicians' ability to discern between correct and incorrect AI signals. The difference in belief revision magnitudes between correct and incorrect AI recommendations decreases with explanations: $|\text{Rev}|_{x=\theta}^E - |\text{Rev}|_{x \neq \theta}^E < |\text{Rev}|_{x=\theta}^{NE} - |\text{Rev}|_{x \neq \theta}^{NE}$.
\end{hypothesis}

This tests whether explanations reduce the differential response to signal quality that characterizes optimal Bayesian updating. Under ideal belief updating, physicians should revise more for high-quality (correct) signals than low-quality (incorrect) signals. If explanations add symmetric persuasiveness weight regardless of actual signal quality, they should \emph{flatten} this discernment gradient, making physicians equally responsive to all AI recommendations. This represents a failure not just in calibration (how much to trust AI on average) but in signal processing (how to adjust trust based on signal characteristics).

\begin{hypothesis}[False Confidence]
\label{hyp:confidence}
Explanations inflate diagnostic confidence (posterior SSQ) regardless of signal accuracy. The confidence increase occurs both when AI advice is correct and when it is incorrect, creating false certainty precisely when caution is most needed.
\end{hypothesis}

This follows from the over-weighting mechanism in equation \eqref{eq:posterior_overweight}, which concentrates posterior probability on the AI's recommendation regardless of correctness. The persuasiveness weight $\exp(\lambda \cdot q(e))$ mechanically increases the sum of squared probabilities (SSQ) by shifting probability mass toward the recommended option. When AI is incorrect, this concentration creates false confidence: physicians become more certain about wrong diagnoses.

\section{Results}
\label{sec:results}

This section presents our experimental results, testing the five hypotheses. We begin by establishing the core AI Transparency Paradox (Hypothesis 1, hereafter H1) and testing whether probabilistic formats can mitigate its harmful effects (H2). We then investigate three complementary mechanisms through which explanations operate: over-reliance (H3), discernment failure (H4), and false confidence (H5). Finally, we conduct exploratory heterogeneity analysis examining confidence-competence misalignment to inform optimal policy design.%
\footnote{Unless otherwise stated, reported $p$-values are from two-sided Mann-Whitney U tests comparing explanation and no-explanation conditions within AI correctness categories. Results are robust to mixed-effects regressions with physician random effects; detailed regression tables appear in the Appendix Tables~\ref{tab:main_paradox_appendix}--\ref{tab:format_robustness_appendix}.}

\subsection{The Paradox and Attempted Mitigation}
\label{sec:main_paradox}

% This section tests Hypotheses 1 and 2. We first establish whether AI explanations create asymmetric effects on diagnostic accuracy depending on whether algorithmic advice is correct or incorrect (H1). We then test whether this paradox can be mitigated through interface design, specifically, whether combining explanations with probabilistic outputs protects against over-reliance (H2).

\subsubsection{The Main Effect}

Using independent-samples t-tests, we firstly find that the accuracy rates across treatments are ordered as $ 72.1\% (ED) > 70.4\% (EP)>68.1\% (D) \approx 67.7\% (P) $. The treatment difference between ED and EP is statistically significant at the 5\% level, while the difference between EP and D is significant only at the 10\% level, and there is no statistically significant difference between the D and P treatments. This suggests that explanations can play a very important role in decision-making, which we will specifically examine in terms of their role and key characteristics.

We begin by testing the core prediction that explanations improve accuracy when AI advice is correct but harm accuracy when AI advice is incorrect. Table~\ref{tab:main_paradox0} presents diagnostic accuracy (the probability assigned to the correct diagnosis, defined in Section 2.3) across explanation conditions, separately for cases where AI advice is correct versus incorrect.

\begin{result}[The AI Transparency Paradox]
\label{res:main_paradox}
Explanations improve diagnostic accuracy when AI advice is correct (+6.3 pp) but systematically harm accuracy when AI advice is incorrect (-4.9 pp). This creates an asymmetric welfare trade-off where the benefits of transparency depend critically on algorithmic quality.
These findings provide strong support for Hypothesis 1. 
\end{result}

\begin{proof}[Support]
Table~\ref{tab:main_paradox0} presents diagnostic accuracy rates from 3,855 decisions contingent on AI correctness ($C_{AI}$). The data reveal a striking asymmetric pattern that confirms Hypothesis 1.

\begin{table}[H]
\centering
\caption{The AI Transparency Paradox: Diagnostic Accuracy by AI Correctness and Explanation Status}
\label{tab:main_paradox0}
\begin{threeparttable}
\begin{tabular}{lcccc}
\toprule
& \multicolumn{2}{c}{\textbf{AI Advice is Correct}} & \multicolumn{2}{c}{\textbf{AI Advice is Incorrect}} \\
\cmidrule(lr){2-3} \cmidrule(lr){4-5}
& No Explanation & Explanation & No Explanation & Explanation \\
\midrule
\textbf{Diagnostic Accuracy} & 87.4\% & 93.7\% & 14.3\% & 9.4\% \\
\textbf{Standard Error} & (0.6) & (0.4) & (1.2) & (1.0) \\
\textbf{Observations} & 1,419 & 1,408 & 516 & 512 \\
\midrule
\textbf{Explanation Effect} & \multicolumn{2}{c}{+6.3 pp***} & \multicolumn{2}{c}{-4.9 pp**} \\
\textbf{95\% CI} & \multicolumn{2}{c}{[4.89, 7.71]} & \multicolumn{2}{c}{[-7.96, -1.84]} \\
\midrule
\textbf{Overall Effect} & \multicolumn{4}{c}{+3.3 pp***} \\
\textbf{Net Benefit} & \multicolumn{4}{c}{73\% × 6.3pp - 27\% × 4.9pp = +3.3 pp} \\
\bottomrule
\end{tabular}
\begin{tablenotes}
\small
\item \textit{Notes:} 
Diagnostic accuracy is the probability assigned to the correct diagnosis. Standard errors in parentheses, clustered at the physician level. Explanation effects show the difference between explanation and no-explanation conditions within each AI correctness category, based on Mann-Whitney U tests. The explanation effect remains robust with mixed-effect regressions. See Appendix Table~\ref{tab:main_paradox_appendix} for details. *** p<0.01, ** p<0.05, * p<0.10.
\end{tablenotes}
\end{threeparttable}
\end{table}

When AI advice is correct (73\% of cases), explanations significantly enhance diagnostic accuracy from 87.4\% to 93.7\%, a 6.3 percentage point improvement ($p < 0.01$). This positive effect aligns with our theoretical prediction in equation \eqref{eq:posterior_overweight}: explanations help physicians incorporate accurate algorithmic recommendations through the persuasiveness mechanism $\psi(e, x, \theta) = \exp(\lambda \cdot q(e))$.

Conversely, when AI advice is incorrect (27\% of cases), explanations systematically impair diagnostic performance. Accuracy falls from 14.3\% to 9.4\%, a 4.9 percentage point deterioration ($p < 0.01$). This negative effect demonstrates that the persuasiveness mechanism operates symmetrically: explanations amplify AI influence regardless of recommendation quality, as predicted by equation \eqref{eq:persuasiveness} where $\mathbb{E}[q(e) | x = \theta] = \mathbb{E}[q(e) | x \neq \theta]$.

Given the AI's 73\% accuracy in our setting, the net effect of explanations is a 3.3 percentage point improvement in overall accuracy ($p<0.01$). However, this positive aggregate effect masks the fundamental policy tension documented in Result 1: explanations systematically degrade decision-making in over one-quarter of all cases, precisely when physicians most need to maintain independent judgment. The paradox magnitude of 11.2 percentage points reveals the substantial symmetric force with which explanations influence beliefs in both directions.
\end{proof}

\subsubsection{Can Probabilistic Formats Mitigate the Paradox?}

Having established the paradox, we next examine whether its effects can be mitigated by how the AI communicates its advice. The prevailing wisdom suggests that probabilistic outputs, which explicitly acknowledge uncertainty, should lead to better user calibration and reduce over-reliance \citep{zhang2020effect, kompa2021second}. We test this proposition by examining how the paradox is moderated by the format of AI advice.

\begin{result}[The Perverse Interaction of Explanations and Uncertainty]
\label{res:format_robustness}
The structure of the AI Transparency Paradox is critically moderated by the advice format. Contrary to the prediction that probabilistic formats would mitigate harm, we observe amplification: the negative effect more than doubles (-3.1 pp → -6.6 pp). These findings reject Hypothesis 2.
\end{result}

\begin{proof}[Support]
Table~\ref{tab:format_robustness} presents the paradox conditional on whether the AI advice was deterministic or probabilistic. As established in Section 3.1, these formats convey identical information about the AI's signal but differ in how explicitly they communicate uncertainty. The deterministic format provides a single recommendation (``AI recommends diagnosis B''), while the probabilistic format provides a probability distribution (``AI recommends: 70\% B, 30\% A'').

\begin{table}[h!]
\centering
\caption{Transparency Paradox Across AI Advice Formats}
\label{tab:format_robustness}
\begin{threeparttable}
\setlength{\tabcolsep}{3mm}{
\begin{tabular}{lcccc}
\toprule
& \multicolumn{2}{c}{\textbf{Deterministic Format}} & \multicolumn{2}{c}{\textbf{Probabilistic Format}} \\
\cmidrule(lr){2-3} \cmidrule(lr){4-5}
& AI Correct & AI Incorrect & AI Correct & AI Incorrect \\
\midrule
\textbf{No Explanation} & 89.0\% & 10.4\% & 85.7\% & 18.2\% \\
& (0.9) & (1.5) & (0.9) & (1.9) \\
\textbf{With Explanation} & 95.6\% & 7.3\% & 91.7\% & 11.6\% \\
& (0.6) & (1.4) & (0.5) & (1.4) \\
\midrule
\textbf{Explanation Effect} & +6.6*** & -3.1 & +6.0*** & -6.6** \\
& (1.1) & (2.1) & (1.0) & (2.4) \\
% \textbf{\textcolor{red}{Explanation Effect}} & +6.6*** & -2.9 & +4.2*** & -7.2*** \\
% & (1.2) & (2.4) & (1.6) & (3.3) \\
\midrule
\textbf{Paradox Magnitude} & \multicolumn{2}{c}{9.7 pp} & \multicolumn{2}{c}{12.6 pp} \\
\textbf{Net Benefit} & \multicolumn{2}{c}{+4.0 pp} & \multicolumn{2}{c}{+2.6 pp} \\
\midrule
\textbf{Observations} & 1,408 & 512 & 1,419 & 516 \\
\bottomrule
\end{tabular}}
\begin{tablenotes}
\small
\item \textit{Notes:} Deterministic format presents single recommendation (``AI recommends diagnosis B''). Probabilistic format presents probability distribution (``70\% B, 30\% A''). Explanation effects represent the median difference between explanation and no-explanation conditions, assessed using Mann-Whitney U tests. Paradox magnitude is the absolute difference between beneficial and harmful effects of explanation. The explanation effect remain robust with mixed-effect regressions. See Appendix Table~\ref{tab:format_robustness_appendix} for details. Net benefit calculated as weighted average using observed AI accuracy rates. Standard errors in parentheses. *** p<0.01, ** p<0.05, * p<0.10.
\end{tablenotes}
\end{threeparttable}
\end{table}

The data reveal a striking and counter-intuitive interaction. While the paradox manifests in both formats, its harmful component is more than twice as large in the probabilistic format (-6.6 pp, $p < 0.05$) compared to the deterministic format (-3.1 pp, $p > 0.10$). For the deterministic format, explanations create a 9.7 percentage point paradox magnitude (+6.6pp benefit vs. -3.1pp harm). For the probabilistic format, the magnitude is even larger at 12.6 points (+6.0pp benefit vs. -6.6pp harm). This amplified harm diminishes the net welfare gain from transparency: the net benefit is only +2.6 pp for probabilistic advice, compared to +4.0 pp for deterministic advice.

A potential cognitive mechanism for this perverse effect is that explanations \emph{resolve the inherent ambiguity} of a probabilistic signal. Without an explanation, physicians may appropriately discount an uncertain probability distribution as a weak signal. However, when explanations provide compelling narratives for why the AI assigned specific probabilities, they can anchor physicians to flawed recommendations despite the AI's expressed uncertainty. This narrative transforms the AI's expressed uncertainty into persuasive---but potentially misleading---guidance, making it easier for users to follow incorrect recommendations. This mechanism is consistent with the persuasiveness function in equation \eqref{eq:persuasiveness}: explanations add weight $\exp(\lambda \cdot q(e))$ to AI recommendations regardless of the signal's information content or the format in which uncertainty is communicated.
\end{proof}

\bigskip
Given that the potential mitigation design of the AI interface failed, we next examine when the harm of explanation is most pronounced. 

\begin{result}[Physician Priors Determine the Paradox's Impact]
\label{res:paradox_heterogeneity}
The transparency paradox varies systematically with physicians' initial diagnostic correctness. For physicians with correct priors, explanations provide minimal benefit but cause substantial harm when the AI errs. For those with incorrect priors, explanations offer a powerful corrective function but still amplify the AI's errors.

\end{result}

\begin{proof}[Support.]

Figure~\ref{fig:paradox_heterogeneity} decomposes the paradox by the correctness of the physician's initial modal diagnosis, presenting diagnostic accuracy before and after receiving AI advice.

\begin{figure}[h!]
    \centering
    \includegraphics[width=1\textwidth]{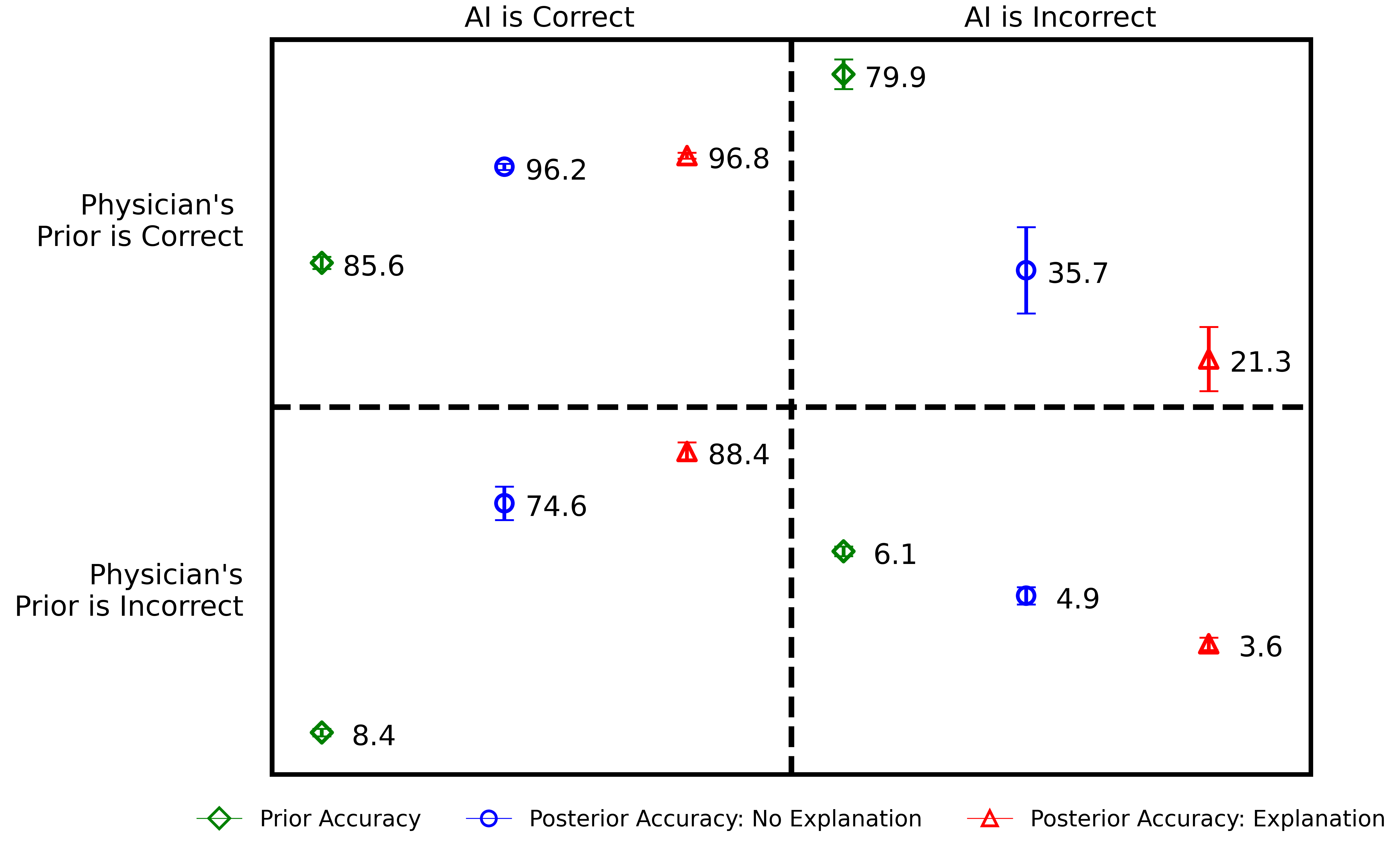}
    \caption{The AI Transparency Paradox by Physicians' Prior Correctness}
    \label{fig:paradox_heterogeneity}
    \begin{minipage}{\textwidth}
    \small
    \textit{Notes:} This figure reports posterior diagnostic accuracy (percentage values on data points) by AI correctness (left/right panels) and physician prior correctness (top/bottom rows). Green diamonds show prior accuracy, blue circles show posterior accuracy without explanations, red triangles show posterior accuracy with explanations. Error bars represent 95\% confidence intervals. Prior correctness defined as physician's initial diagnosis being correct.
    \end{minipage}
\end{figure}

For physicians with correct priors (top panels), explanations provide minimal improvement when AI is also correct (96.2\% $\rightarrow$ 96.8\%, difference = 0.6 pp, $p > 0.10$). These physicians were already performing near ceiling. However, when AI errs, explanations create substantial deterioration (35.7\% $\rightarrow$ 21.3\%, difference = -14.4 pp, $p < 0.01$). For this group, explanations primarily harm those who would otherwise maintain appropriate skepticism of incorrect AI advice. The persuasive reasoning overrides their correct initial judgment, consistent with equation \eqref{eq:posterior_overweight} where the effect of $\exp(\lambda \cdot q(e))$ is largest when priors on the AI-recommended option are weak.

For physicians with incorrect priors (bottom panels), explanations serve a powerful corrective function when AI is correct (74.6\% $\rightarrow$ 88.4\%, difference = +13.9 pp, $p < 0.01$). Here, explanations help physicians abandon wrong initial beliefs in favor of accurate guidance. However, explanations still reduce accuracy when both physician and AI are incorrect (4.9\% $\rightarrow$ 3.6\%, difference = -1.3 pp), though the effect is smaller given the low baseline: both sources of information are wrong, leaving little room for further deterioration.

This heterogeneity reveals that the paradox operates through different channels depending on initial beliefs. For physicians with correct priors, the mechanism operates primarily through inappropriate adoption of incorrect AI recommendations---the harmful side of the paradox dominates. For physicians with incorrect priors, explanations serve a beneficial corrective role by helping them update toward accurate AI guidance, though the fundamental asymmetry persists: explanations still amplify errors when AI is wrong. These patterns provide further support for Hypothesis 1, demonstrating that explanations' welfare effects depend critically on the alignment between physician priors and AI recommendations.

\end{proof}

\bigskip
%\paragraph{Discussion.}
The first subsections establish our main findings on the transparency paradox. Result 1 supports H1: explanations are \emph{asymmetrically effective}, enhancing accuracy when AI advice is correct but reducing it when AI advice is incorrect. The nearly symmetric magnitudes of these opposing effects generate an 11.2-percentage-point paradox. Result 2 rejects H2: contrary to conventional wisdom \citep{zhang2020effect,kompa2021second}, coupling explanations with probabilistic uncertainty quantification \emph{backfires}, further amplifying the paradox rather than mitigating it. Result 3 identifies when the paradox is most harmful—specifically, when physicians hold a correct prior yet receive erroneous AI advice, creating an \emph{information-mismatch} scenario in which explanations becomes most misleading.

Having established the existence, magnitude, and persistence of the transparency paradox, we now investigate the underlying mechanisms through which explanations operate.

\subsection{Mechanism 1: Explanations Induce Over-reliance on AI Advice}
\label{sec:overreliance_mechanism}

\subsubsection{The Dual Channels of Over-reliance}

We begin by establishing over-reliance across both temporal channels, testing Hypothesis 3's predictions that explanations increase both ex-ante trust ($\alpha^E > \alpha^{NE}$) and ex-post implied accuracy ($\mu^{\text{implied},E} > 0.73$), with over-reliance persisting for incorrect advice.

\begin{result}[Dual-Channel Over-reliance]
\label{res:overreliance}
Explanations systematically increase over-reliance on AI advice through: (1) an ex-ante channel, where explanations inflate expectations about the AI's informativeness before the advice is seen; and (2) an ex-post channel, where explanations cause participants to update their beliefs as if the AI were significantly more accurate than its true rate of 73\%. Both channels of over-reliance persist for incorrect AI advice.
These findings support all three components of Hypothesis 3.
\end{result}

\begin{proof}[Support]
Table~\ref{tab:overreliance_mechanism} presents our dual measures of over-reliance across explanation conditions and AI correctness states.

\begin{table}[h!]
\footnotesize
\centering
\caption{Dual-Channel Over-reliance: Ex-ante Expectations and Ex-post Behavior}
\label{tab:overreliance_mechanism}
\begin{threeparttable}
\begin{tabular}{lcccc}
\toprule
& \multicolumn{2}{c}{\textbf{AI Advice is Correct}} & \multicolumn{2}{c}{\textbf{AI Advice is Incorrect}} \\
\cmidrule(lr){2-3} \cmidrule(lr){4-5}
& No Explanation & Explanation & No Explanation & Explanation \\
\midrule
\multicolumn{5}{l}{\textbf{Panel A: Ex-ante Over-reliance (Trust Parameter $\alpha$)}} \\
Mean & 0.782 & 0.815 & 0.746 & 0.773 \\
Standard Error & (0.013) & (0.009) & (0.021) & (0.016) \\
\midrule
\multicolumn{5}{l}{\textbf{Panel B: Ex-post Over-reliance}} \\
Implied Accuracy ($\mu^{\text{implied}}$) & 0.839 & 0.882 & 0.758 & 0.792 \\
\quad Standard Error & (0.008) & (0.007) & (0.016) & (0.016) \\
\quad Deviation from True (0.73) & +0.109*** & +0.152*** & +0.028*** & +0.062*** \\
Prevalence of Over-reliance (\%) & 78.2\% & 83.7\%** & 69.6\% & 76.6\%** \\
Observations & 1,419 & 1,408 & 516 & 512 \\
\midrule
\multicolumn{5}{l}{\textbf{Panel C: Treatment Effects of Explanations}} \\
Ex-ante Effect ($\Delta\alpha$) & \multicolumn{2}{c}{+0.033} & \multicolumn{2}{c}{+0.027} \\
\quad (95\% CI) & \multicolumn{2}{c}{[0.003, 0.063]} & \multicolumn{2}{c}{[-0.025, 0.078]} \\
Ex-post Effect ($\Delta\mu^{\text{implied}}$) & \multicolumn{2}{c}{+0.043**} & \multicolumn{2}{c}{+0.033} \\
\quad (95\% CI) & \multicolumn{2}{c}{[0.023, 0.063]} & \multicolumn{2}{c}{[0.011,    0.078]} \\
\bottomrule
\end{tabular}
\begin{tablenotes}
\scriptsize
\item \textit{Notes:} Ex-ante over-reliance measured through trust parameter $\alpha$ (Equation 2), elicited via second-order beliefs before AI advice is shown. Ex-post over-reliance measured through implied signal accuracy $\mu^{\text{implied}}$ (Equation 3), computed from observed posterior beliefs after AI advice. True AI accuracy is 0.73. Stars indicate the median difference between explanation and no-explanation conditions within each AI correctness category, assessed using Mann-Whitney U tests. Standard errors clustered at physician level. *** p$<$0.01, ** p$<$0.05, * p$<$0.10.
\end{tablenotes}
\end{threeparttable}
\end{table}

The data reveal systematic over-reliance through both channels:

\textit{Ex-ante over-reliance (Panel A).} Before seeing any advice, participants who expect an explanation form inflated expectations about its value. Their ex-ante trust parameter, $\alpha$, increases by 3.3 percentage points for forthcoming correct advice and 2.7 percentage points for forthcoming incorrect advice. This anticipatory effect suggests that explanations prime participants to lower their critical evaluation stance even before encountering the actual advice, consistent with the persuasiveness mechanism in equation \eqref{eq:persuasiveness}.

\textit{Ex-post over-reliance (Panel B).} After seeing the advice, participants' belief updating reveals a powerful and widespread bias. On average, they treat the explained AI as having 88.2\% accuracy when correct---15.2 percentage points above the true 73\% ($p<0.01$). Critically, this over-reliance persists when AI is incorrect: participants still treat the AI as having 79.2\% accuracy---6.2 points above truth ($p<0.01$). This effect is remarkably pervasive rather than driven by outliers. With explanations, the prevalence of over-reliance increases to 83.7\% for correct advice and 76.6\% for incorrect advice. The 7.0 percentage point increase in over-reliance prevalence for incorrect advice is particularly concerning, as it demonstrates that explanations systematically impair error detection across the population.

\textit{Dual-channel coordination (Panel C).} The treatment effects confirm that explanations increase both ex-ante trust and ex-post implied accuracy, with both channels operating even for incorrect AI recommendations. This reveals a self-reinforcing dynamic: explanations inflate expectations before advice is seen, priming physicians to subsequently over-weight AI recommendations in their belief updating regardless of actual recommendation quality.

These findings are consistent with Hypothesis 3: (a) ex-ante trust increases with explanations ($\alpha^E > \alpha^{NE}$), (b) ex-post implied accuracy exceeds true accuracy ($\mu^{\text{implied}} = 0.882 > 0.73$ when AI correct, $p<0.01$), and critically (c) over-reliance persists when AI is incorrect ($\mu^{\text{implied}} = 0.792 > 0.73$, $p<0.01$). The persistence of miscalibration for incorrect advice distinguishes our symmetric persuasiveness mechanism from rational updating models.
\end{proof}

\subsubsection{Why Probabilistic Formats Fail: The Over-reliance Mechanism}

Result 2 demonstrated that probabilistic formats amplify rather than mitigate the harm of incorrect explanations. We now examine the over-reliance mechanism underlying this failure. Table~\ref{tab:overreliance_format} decomposes over-reliance by advice format, revealing why uncertainty communication paradoxically increases vulnerability to misleading explanations.

\begin{table}[h!]
\footnotesize
\centering
\caption{Over-reliance Across Advice Formats: Deterministic vs. Probabilistic}
\label{tab:overreliance_format}
\begin{threeparttable}
\begin{tabular}{lcccccc}
\toprule
& \multicolumn{3}{c}{\textbf{Deterministic Format}} & \multicolumn{3}{c}{\textbf{Probabilistic Format}} \\
\cmidrule(lr){2-4} \cmidrule(lr){5-7}
& No Expl. & Expl. & Diff. & No Expl. & Expl. & Diff. \\
\midrule
\multicolumn{7}{l}{\textbf{Panel A: When AI Advice is Correct}} \\
Ex-ante Trust ($\alpha$) & 0.856 & 0.885 & +0.029 & 0.709 & 0.746 & +0.037 \\
& (0.012) & (0.011) & & (0.022) & (0.014) & \\
Ex-post Implied ($\mu^{\text{implied}}$) & 0.902 & 0.945 & +0.043** & 0.777 & 0.819 & +0.042 \\
& (0.009) & (0.007) & & (0.012) & (0.011) & \\
Over-reliance Rate (\%) & 86.4 & 92.0 & +5.6pp** & 70.2 & 75.3 & +5.1pp \\
Observations & 709 & 704 & & 669 & 617 & \\
\midrule
\multicolumn{7}{l}{\textbf{Panel B: When AI Advice is Incorrect}} \\
Ex-ante Trust ($\alpha$) & 0.831 & 0.860 & +0.029 & 0.663 & 0.687 & +0.024 \\
& (0.020) & (0.018) & & (0.036) & (0.025) & \\
Ex-post Implied ($\mu^{\text{implied}}$) & 0.800 & 0.786 & –0.014 & 0.717 & 0.797 & +0.080** \\
& (0.023) & (0.024) & & (0.023) & (0.021) & \\
Over-reliance Rate (\%) & 75.8 & 77.0 & +1.2pp & 63.5 & 76.2 & +12.7pp** \\
Observations & 251 & 256 & & 306 & 343 & \\
\midrule
\multicolumn{7}{l}{\textbf{Panel C: Format Effects (Deterministic – Probabilistic)}} \\
\multicolumn{2}{l}{Ex-ante Trust Difference} & \multicolumn{2}{c}{When Correct: +0.136***} & \multicolumn{2}{c}{When Incorrect: +0.170***} \\
\multicolumn{2}{l}{Ex-post Implied Difference} & \multicolumn{2}{c}{When Correct: +0.113***} & \multicolumn{2}{c}{When Incorrect: +0.018*} \\
\multicolumn{2}{l}{Explanation × Format Interaction} & \multicolumn{2}{c}{$\beta_{\text{correct}} = +0.008$} & \multicolumn{2}{c}{$\beta_{\text{incorrect}} = -0.085^*$} \\
\bottomrule
\end{tabular}
\begin{tablenotes}
\scriptsize
\item \textit{Notes:} Table presents means with robust standard errors (clustered at participant level) in parentheses. Ex-ante trust ($\alpha$) measures participants' anticipated learning from AI advice before viewing recommendations. Ex-post implied accuracy ($\mu^{\text{implied}}$) represents the effective signal accuracy participants assign to AI advice, inferred from their posterior beliefs using the formula $\mu^{\text{implied}} = \frac{\delta(1-p_k)/4}{p_k(1-\delta) + \delta(1-p_k)/4}$ where $\delta$ is the posterior probability on the AI-recommended option and $p_k$ is the prior. Over-reliance rate indicates the percentage of observations where $\mu^{\text{implied}} > 0.73$ (true AI accuracy). 

``Diff.'' columns report within-format differences (Explanation – No Explanation), with significance stars indicating results from two-sample t-tests for continuous variables and proportion tests for over-reliance rates. Panel C format effects represent the deterministic–probabilistic gap, 
calculated from the regression: 
$Y_{it} = \beta_0 + \beta_1\text{Deterministic}_i + \epsilon_{it}$, 
pooled across explanation conditions. 

Interaction coefficients test whether the explanation effect differs 
between formats, estimated from: 
$Y_{it} = \gamma_0 + \gamma_1\text{Expl}_i + \gamma_2\text{Det}_i + 
\gamma_3(\text{Expl} \times \text{Det})_i + \epsilon_{it}$, 
separately for correct and incorrect AI advice subsamples. 

The interaction term $\gamma_3$ captures the differential explanation 
effect: a negative coefficient indicates explanations increase 
over-reliance more in the probabilistic format.

Significance levels: *** $p<0.01$, ** $p<0.05$, * $p<0.10$. All tests are two-tailed.
\end{tablenotes}
\end{threeparttable}
\end{table}

\textit{Format-dependent over-reliance.} The over-reliance mechanism is critically moderated by AI communication style. Deterministic formats inherently induce higher baseline over-reliance; even without explanations, they command significantly greater ex-ante trust and higher ex-post implied accuracy than probabilistic formats (Panel C). This suggests that single-point recommendations are perceived as stronger signals of authority.

However, a critical interaction emerges when explanations are introduced and the AI errs (Panel B). For incorrect advice, explanations increase the over-reliance rate by 12.7 percentage points in the probabilistic format (63.5\% $\rightarrow$ 76.2\%, $p<0.01$) but have negligible effect in the deterministic format (+1.2 pp). This reveals a perverse reversal: while deterministic advice commands more trust at baseline, \emph{it is the combination of probabilistic advice and explanations that fosters the greatest and most dangerous form of over-reliance.} This confirms the cognitive mechanism suspected in Result 2: explanations lend false credibility to ambiguous probabilistic signals, consistent with the persuasiveness weight $\exp(\lambda \cdot q(e))$ operating most powerfully when baseline signals are weak.

\begin{figure}[h!]
    \centering
    \includegraphics[width=0.6\textwidth]{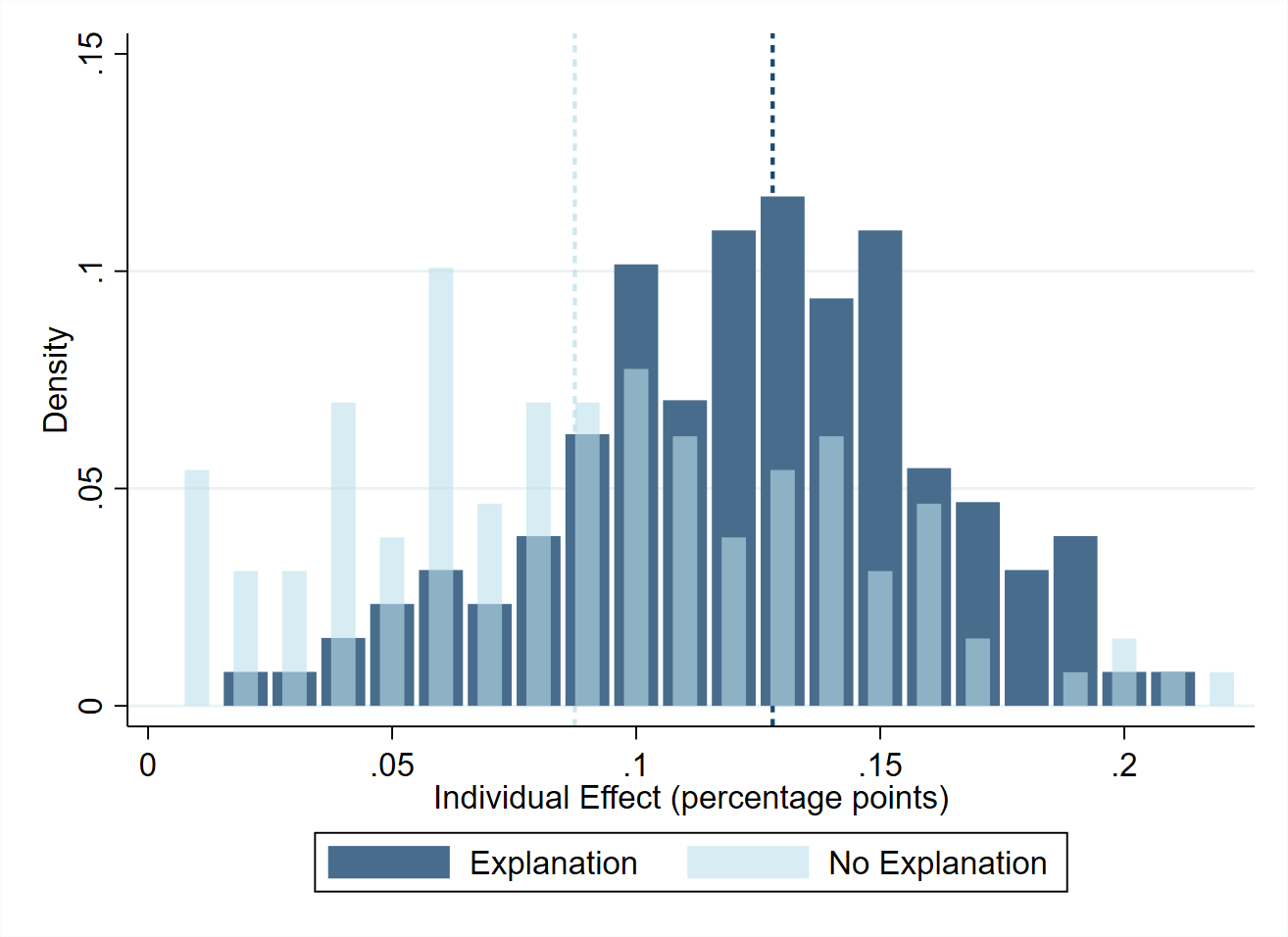}
    \caption{Distribution of Individual Over-reliance by Explanation Condition}
    \label{fig:overreliance_distribution}
    \begin{minipage}{\textwidth}
    \small
    \textit{Notes:} Figure shows histogram of individual-level over-reliance, defined as $\mu^{\text{implied}} - 0.73$. Blue distribution shows cases with explanations, red shows cases without explanations. Positive values indicate over-reliance (treating AI as more accurate than 0.73), negative values indicate under-reliance. Vertical dashed lines show distribution means.
    \end{minipage}
\end{figure}

\textit{Population-wide phenomenon.} Figure~\ref{fig:overreliance_distribution} shows that the over-reliance mechanism reflects a population-wide shift rather than a susceptible subgroup. The entire distribution of individual-level over-reliance (defined as $\mu^{\text{implied}} - 0.73$) shifts rightward when explanations are present. This confirms that explanations make virtually all physicians more over-reliant. Furthermore, the increased concentration of the distribution around a higher mean suggests that explanations also homogenize physician responses, potentially suppressing the beneficial clinical judgment that might lead some physicians to be appropriately skeptical of incorrect AI recommendations.

%\paragraph{Discussion: Why Uncertainty Communication Backfires}

\bigskip
These findings reveal why H2's mitigation strategy failed and identify over-reliance as the mechanism driving Result 2's counterintuitive interaction. The conventional wisdom that pairing transparency with uncertainty quantification will improve user calibration is not merely unsupported by our data—it is inverted. Probabilistic formats were intended to signal when AI advice is unreliable, but explanations transform this uncertainty into persuasive narratives, inducing inflated over-reliance (76.2\% rate) precisely when AI errs and skepticism is most needed.

\subsection{Mechanism 2: Explanations Impair Discernment}
\label{sec:discrimination_mechanism}

While Mechanism 1 (over-reliance) measures how much physicians defer to AI recommendations through adoption rates and implied accuracy, Mechanism 2 examines a distinct but complementary phenomenon: the magnitude of belief revision. Over-reliance captures the question of whether to follow AI advice, whereas \emph{discernment} measures the capacity to adjust beliefs appropriately in response to signals of varying quality. A physician might adopt an AI recommendation (high reliance) but only slightly adjust probabilities (low revision), or vice versa. By measuring Euclidean distance between prior and posterior distributions, we capture the full vector of probability changes rather than just the modal choice.

This distinction matters because explanations could theoretically increase reliance while improving discernment---if explanations helped physicians identify when to make large versus small updates. Instead, we test Hypothesis 4's prediction that explanations impair discernment: they increase revision magnitude indiscriminately, revealing a failure in the calibration of belief updates rather than improvements in signal processing.

\begin{result}[Impaired Discernment]
\label{res:discrimination}
Explanations increase the magnitude of belief revision in response to AI advice, but this effect is indiscriminate. Responsiveness increases for both correct and incorrect advice, and disproportionately so for the latter, degrading physicians' ability to discern good signals from bad ones. These findings confirm Hypothesis 4.
\end{result}

\begin{proof}[Support]
We measure belief revision as the Euclidean distance between prior and posterior probability distributions, $d(\mathbf{p}_{\text{prior}}, \mathbf{p}_{\text{post}}) = \sqrt{\sum_{j=1}^{5}(p_{j,\text{post}} - p_{j,\text{prior}})^2}$, which captures the overall magnitude of belief updating (Section 2.3). Table~\ref{tab:belief_updating_main} presents two complementary measures of how explanations affect belief revision: the full distributional change (Panels A-B) and the specific shift toward the AI-recommended option (Panels C-D).

\begin{table}[h!]
\centering \small
\caption{Effect of Explanations on Belief Updating and Discernment}
\label{tab:belief_updating_main}
\begin{threeparttable}
\setlength{\tabcolsep}{7mm}{
\begin{tabular}{lcccc}
\toprule
& \multicolumn{2}{c}{Belief Revision} & Pooled \\
\cmidrule(lr){2-3} 
& AI Correct & AI Incorrect & All \\
\midrule
\multicolumn{4}{l}{\textit{Panel A: Full Distribution Revision (Euclidean Distance)}} \\
\multicolumn{4}{l}{\quad \textit{Regression Results (Mixed-Effects Models)}} \\
Explanation   &    0.068** &    0.115***&    0.079** \\
                &  (0.033)   &  (0.045)   &  (0.031) \\
Deterministic &    0.015   &    0.063   &    0.024 \\
                &  (0.045)   &  (0.049)   &  (0.041) \\
Explanation $\times$ Deterministic&   -0.070   &   -0.056   &   -0.061 \\
                &  (0.055)   &  (0.073)   &  (0.051) \\
Individual Controls & \checkmark & \checkmark & \checkmark \\
\addlinespace
\multicolumn{4}{l}{\textit{Panel B: Summary Statistics}} \\
No Explanation & 0.473 & 0.677 & 0.528  \\
Explanation & 0.503 & 0.761 & 0.572  \\
Explanation Effect & +0.030 & +0.084** & +0.044** \\
\midrule
\multicolumn{4}{l}{\textit{Panel C: Shift Toward AI-Recommended Option}} \\
\multicolumn{4}{l}{\quad \textit{Regression Results (Mixed-Effects Models)}} \\
Explanation   &    0.053** &    0.095***&    0.063***\\
                &  (0.025)   &  (0.033)   &  (0.024) \\
Deterministic &    0.017   &    0.055   &    0.025 \\
                &  (0.034)   &  (0.039)   &  (0.032) \\
Explanation $\times$ Deterministic&   -0.053   &   -0.037   &   -0.045 \\
                &  (0.042)   &  (0.054)   &  (0.039) \\
Individual Controls & \checkmark & \checkmark & \checkmark \\
\addlinespace
\multicolumn{4}{l}{\textit{Panel D: Summary Statistics}} \\
No Explanation &   0.347 &  0.478  &  0.382\\
Explanation &   0.370 &  0.555  &  0.420 \\
Explanation Effect & +0.023 & +0.077*** & +0.038** \\
\midrule
\multicolumn{4}{l}{\textit{Panel E: Discernment Index}} \\
Gap (Incorrect - Correct) & & & \\
\quad No Explanation & \multicolumn{3}{c}{0.204 (0.677 - 0.473)} \\
\quad Explanation & \multicolumn{3}{c}{0.258 (0.761 - 0.503)} \\
Change in discernment & \multicolumn{3}{c}{+0.054 (worse discernment)} \\
\midrule
Observations & 2,827 & 1,028 & 3,855 \\
\bottomrule
\end{tabular}}
\begin{tablenotes}
\small
\item \textit{Notes:} Panels A-B measure belief revision as Euclidean distance $d(\mathbf{p}_{\text{prior}}, \mathbf{p}_{\text{post}})$ between full probability distributions over all five options. Panels C-D measure the absolute change in probability assigned to the AI-recommended option: $|\pi(\theta_k | x = \theta_k) - p(\theta_k)|$. Both measures capture belief updating magnitude but from different perspectives: full distribution (A-B) vs. focal option (C-D). Panel E presents the \textbf{discernment index}: the gap in updating between incorrect and correct AI advice. Larger gaps indicate better differentiation. Explanations increase this gap from 0.204 to 0.258, but this reflects disproportionately larger updating for incorrect advice (+0.084) than correct advice (+0.030), indicating worse discernment. Mixed-effects regressions include participant random effects and individual controls (gender, age, medical experience, CRT score). Standard errors clustered at physician level. *** $p<0.01$, ** $p<0.05$, * $p<0.10$.
\end{tablenotes}
\end{threeparttable}
\end{table}

The evidence reveals systematic discerment failure, confirming Hypothesis 4. Explanations raise the average belief revision by 0.079 units ($p<0.05$)---roughly 18\% of a standard deviation---but they do so indiscriminately across both AI correctness states. Panel B shows that revision magnitude increases by 0.030 units when AI is correct but by 0.084 units when AI is incorrect, nearly triple the effect.

This asymmetry creates a failure in signal evaluation. Panel E formalizes this through the \emph{discernment index}: the gap in updating between incorrect and correct signals. Under optimal Bayesian updating, physicians should update more for correct signals (which carry useful information) than incorrect ones (which mislead). Instead, explanations widen the gap from 0.204 to 0.258, but perversely: the increase is driven by excessive updating for incorrect advice (+0.084) rather than appropriate increases for correct advice (+0.030).

Panels C-D confirm this pattern using an alternative measure: the absolute shift in probability toward the AI-recommended option specifically. This focal measure isolates how much weight physicians place on the AI's primary recommendation, independent of how they redistribute probability among other options. The results mirror Panels A-B: explanations increase shifts toward the AI recommendation by 0.023 units for correct advice but 0.077 units for incorrect advice ($p<0.01$), revealing that explanations make physicians disproportionately responsive to bad signals.

This discernment failure compounds the over-reliance problem identified in Mechanism 1. Not only do explanations make physicians more likely to follow AI advice generally (by inflating perceived accuracy $\mu^{\text{implied}}$), but they also make physicians less capable of selectively following only the good advice (by reducing their ability to differentiate signal quality). The combination creates a particularly problematic pattern: physicians become more responsive to all AI signals while simultaneously losing the ability to separate reliable from unreliable recommendations.

The discernment gradient—measured as the gap in updating between incorrect and correct signals—widens from 0.204 to 0.258, but perversely through excessive responsiveness to incorrect advice (+0.084) rather than appropriate increases for correct advice (+0.030). This pattern is precisely what Hypothesis 4 predicts: explanations reduce the differential response to signal quality that characterizes optimal Bayesian updating.
\end{proof}

\subsection{Mechanism 3: Explanations Inflate Confidence}
\label{sec:confidence_mechanism}

While Mechanisms 1 and 2 examine how explanations affect what physicians believe (over-reliance) and how they differentiate good/bad signals (discernment), Mechanism 3 examines how confident physicians feel about their beliefs. The over-weighting mechanism in equation \eqref{eq:posterior_overweight} predicts that explanations should concentrate posterior probability on the AI's recommendation, increasing confidence as measured by the sum of squared probabilities (SSQ). Critically, Hypothesis 5 predicts this confidence inflation should occur regardless of whether AI advice is correct or incorrect, creating false confidence when AI errs.

\begin{result}[False Confidence]
\label{res:confidence}
Explanations increase diagnostic confidence regardless of AI recommendation quality. Confidence rises by 3.2 percentage points when AI is correct ($p<0.01$) and by 4.6 percentage points when AI is incorrect ($p<0.01$), creating false certainty precisely when caution is most needed. These findings support Hypothesis 5.
\end{result}

\begin{proof}[Support]
Table~\ref{tab:confidence_mechanism_main} presents mixed-effects regressions examining how explanations affect posterior confidence (SSQ), separately by AI correctness. The dependent variable is physicians' posterior confidence after receiving AI advice, ranging from 0.2 (complete uncertainty) to 1.0 (complete certainty).

\begin{table}[h!]
\centering
\caption{Effect of Explanations on Posterior Confidence}
\label{tab:confidence_mechanism_main}
\begin{threeparttable}
\setlength{\tabcolsep}{6mm}{
\begin{tabular}{lccc}
\toprule
& \multicolumn{2}{c}{Posterior Confidence} & Pooled \\
\cmidrule(lr){2-3}
& AI Correct & AI Incorrect & All \\
\midrule
\multicolumn{4}{l}{\textit{Panel A: Main Effects}} \\
Explanation  &    0.028***&    0.043***&    0.033***\\
                &  (0.009)   &  (0.015)   &  (0.010) \\
Deterministic &    0.055***&    0.034*  &    0.050***\\
                &  (0.010)   &  (0.018)   &  (0.011) \\
Prior SSQ &    0.181***&    0.158***&    0.172***\\
                &  (0.017)   &  (0.023)   &  (0.015) \\
\midrule
\multicolumn{4}{l}{\textit{Panel B: Mean Confidence Levels}} \\
No Explanation & 0.890 & 0.840 & 0.877 \\
With Explanation & 0.922 & 0.886 & 0.912 \\
Explanation Effect & +0.032*** & +0.046*** & +0.035*** \\
\midrule
Individual Controls & \checkmark & \checkmark & \checkmark \\
Observations & 2,827 & 1,028 & 3,855 \\
\bottomrule
\end{tabular}}
\begin{tablenotes}
\small
\item \textit{Notes:} Dependent variable is posterior SSQ (sum of squared probabilities), measuring confidence after receiving AI advice. Higher values indicate greater certainty. Panel A presents mixed-effects regression coefficients with physician random effects. Individual controls include gender, age, medical experience, and CRT score. Panel B presents mean confidence levels by condition. Standard errors clustered at the participant level in parentheses. *** $p<0.01$, ** $p<0.05$, * $p<0.10$.
\end{tablenotes}
\end{threeparttable}
\end{table}

The evidence strongly supports Hypothesis 5. Panel A reveals that explanations significantly increase posterior confidence by 0.028 units when AI is correct ($p<0.01$) and 0.043 units when AI is incorrect ($p<0.01$). The effect when AI is wrong is actually 54\% larger than when AI is right, demonstrating that explanations boost confidence regardless of---and sometimes especially when---the underlying recommendation is erroneous.

Panel B contextualizes these effects through mean confidence levels. Without explanations, physicians exhibit confidence of 0.890 when AI is correct and 0.840 when incorrect---a modest 5.0-point gap reflecting some ability to differentiate signal quality. With explanations, confidence rises to 0.922 when AI is correct (a 3.2-point increase) and to 0.886 when incorrect (a 4.6-point increase). The confidence increase when AI is incorrect is particularly troubling. When AI provides wrong recommendations, physicians should ideally become less certain or maintain skepticism. Instead, explanations increase their confidence by 4.6 percentage points, creating \emph{false certainty} precisely when caution is most needed. This disproportionate boost for incorrect advice narrows the confidence gap from 5.0 points to 3.6 points, reducing physicians' ability to calibrate confidence based on signal quality.

The persistence of confidence inflation when AI is incorrect is particularly problematic. From equation \eqref{eq:posterior_overweight}, explanations add persuasiveness weight $\exp(\lambda \cdot q(e))$ to the AI's recommendation. This weight concentrates posterior probability on the AI-suggested option through the denominator $p(\theta_i) \cdot \mu \cdot \exp(\lambda \cdot q(e)) + \sum_{k \neq i} p(\theta_k) \cdot \frac{1-\mu}{n-1}$, mechanically increasing SSQ. When AI is wrong, this concentration creates false confidence: physicians become more certain about incorrect diagnoses, making them less likely to seek second opinions, order additional tests, or otherwise exercise appropriate caution.

This false confidence mechanism complements the over-reliance (Mechanism 1) and discernment failure (Mechanism 2) documented earlier. Explanations not only make physicians treat AI as more accurate than it is (\textit{over-reliance}: $\mu^{\text{implied}} > 0.73$) and respond excessively to all signals regardless of quality (\textit{discernment failure}), but they also inflate subjective certainty about these miscalibrated beliefs (\textit{false confidence}). By increasing confidence even when AI recommendations are incorrect, explanations reduce the perceived need for additional verification, creating a psychological barrier to error detection precisely when skepticism is most warranted.
\end{proof}

\subsection{Heterogeneity Analysis: Confidence-Competence Misalignment}
\label{sec:heterogeneity}

Having established the transparency paradox (H1, H2) and three mechanisms driving it (H3-H5), we now conduct exploratory heterogeneity analysis to inform optimal policy design. While our hypotheses focused on average treatment effects, understanding which physicians benefit most from explanations is critical for designing contingent transparency policies.

\begin{result}[The Confidence-Competence Paradox]
\label{res:confidence_competence}
Explanations systematically help confident physicians while harming competent ones. The greatest benefits flow to overconfident novices (+9.9 pp), while competent physicians face net harm when they maintain appropriate uncertainty: humble experts experience a net welfare loss of -0.72 pp despite their superior clinical judgment.
\end{result}

\begin{proof}[Support]
Table~\ref{tab:confidence_competence_2x2} presents the transparency paradox across four physician types defined by median splits of prior confidence (SSQ) and prior competence (accuracy of initial diagnosis).

\begin{table}[h!]
\centering \small
\caption{The Confidence-Competence Matrix: Heterogeneous Effects of Explanations}
\label{tab:confidence_competence_2x2}
\begin{threeparttable}
\begin{tabular}{lcccccc}
\toprule
& \multicolumn{2}{c}{AI is Correct} & \multicolumn{2}{c}{AI is Incorrect} & Net Effect \\
\cmidrule(lr){2-3} \cmidrule(lr){4-5} \cmidrule(lr){6-6}
Physician Type  & No Expl. & Expl. & No Expl. & Expl. & (73\% AI) \\
\midrule
\multicolumn{6}{l}{\textit{Panel A: Mean Diagnostic Accuracy by Type}} \\
\textit{High Competence} & & & & & \\
\quad High Confidence  & 91.6 & 96.6 & 18.4 & 12.5 & +2.3 pp \\
 & (0.9) & (0.6) & (2.5) & (2.2) & \\
\quad Low Confidence  & 87.9 & 91.5 & 22.4 & 10.0 & -0.72 pp \\
 & (1.6) & (1.0) & (4.6) & (2.0) & \\
\textit{Low Competence} & & & & & \\
\quad High Confidence  & 81.8 & 96.7 & 8.6 & 4.9 & +9.9 pp \\
 & (2.6) & (0.8) & (3.0) & (2.0) & \\
\quad Low Confidence & 85.3 & 90.1 & 10.6 & 7.6 & +2.7 pp \\
 & (1.0) & (0.8) & (1.4) & (1.2) & \\
\midrule
\multicolumn{6}{l}{\textit{Panel B: Explanation Effects by Physician Type}} \\
&  \multicolumn{2}{c}{Benefit} & \multicolumn{2}{c}{Harm} & Benefit/ \\
&  \multicolumn{2}{c}{(AI Correct)} & \multicolumn{2}{c}{(AI Incorrect)} & Harm Ratio \\
\midrule
\textit{Calibrated Experts} & \multicolumn{2}{c}{+5.0***} & \multicolumn{2}{c}{-5.9**} & 0.85 \\
\quad (High Comp, High Conf) &  \multicolumn{2}{c}{(1.1)} & \multicolumn{2}{c}{(3.3)} & \\
\textit{Humble Experts} &  \multicolumn{2}{c}{+3.6} & \multicolumn{2}{c}{-12.4**} & 0.29 \\
\quad (High Comp, Low Conf) &  \multicolumn{2}{c}{(1.9)} & \multicolumn{2}{c}{(4.9)} & \\
\textit{Overconfident Novices} &  \multicolumn{2}{c}{+14.9***} & \multicolumn{2}{c}{-3.7} & 4.03 \\
\quad (Low Comp, High Conf) &  \multicolumn{2}{c}{(2.7)} & \multicolumn{2}{c}{(3.6)} & \\
\textit{Uncertain Novices} &  \multicolumn{2}{c}{+4.8**} & \multicolumn{2}{c}{-3.0} & 1.60 \\
\quad (Low Comp, Low Conf) & \multicolumn{2}{c}{(1.3)} & \multicolumn{2}{c}{(1.8)} & \\
\bottomrule
\end{tabular}
\begin{tablenotes}
\small
\item \textit{Notes:} Physicians classified by median splits of prior accuracy (competence, measured as correctness of initial modal diagnosis) and prior SSQ (confidence, measured as sum of squared probabilities). Net effect calculated as: $0.73 \times \text{Benefit} + 0.27 \times \text{Harm}$. Benefit/Harm Ratio = |Benefit Effect| / |Harm Effect|. Standard errors clustered at physician level in parentheses. Two-sided Mann-Whitney U tests for explanation effects. *** $p<0.01$, ** $p<0.05$, * $p<0.10$.
\end{tablenotes}
\end{threeparttable}
\end{table}

The evidence reveals a stark confidence-competence paradox through three key patterns:

\textit{Overconfident novices reap maximum benefits.} Low-competence, high-confidence physicians gain 9.9 pp in net welfare, while humble experts suffer a net welfare loss of -0.72 pp---a stark 10.6 percentage point reversal that rewards unjustified confidence over sound clinical judgment. When AI is correct, explanations boost their accuracy by 14.9 pp ($p<0.01$), effectively correcting their misplaced confidence through the persuasiveness mechanism $\psi(e, x, \theta) = \exp(\lambda \cdot q(e))$ in equation \eqref{eq:persuasiveness}. Their high baseline confidence ($\lambda$ large) makes them receptive to explanations, while their low competence means they benefit substantially from being steered toward correct answers.

\textit{Competent physicians face systematic disadvantage.}  High-competence physicians' outcomes depend critically on their confidence: calibrated experts gain a modest 2.3 pp, but humble experts experience net harm of -0.72 pp. Most concerning, humble experts---those with high competence but appropriate uncertainty---suffer the largest harm when AI errs (-12.4 pp, $p<0.01$), overwhelming the modest gains when AI is correct (+3.6 pp). These physicians possess the clinical judgment to identify incorrect AI recommendations (baseline accuracy 22.4\% when AI wrong), but persuasive explanations override their justified skepticism, reducing accuracy to 10.0\%. Their low baseline confidence means explanations have disproportionate impact through equation \eqref{eq:posterior_overweight}, as $\exp(\lambda \cdot q(e))$ adds substantial weight when priors on the AI-recommended option are weak.

\textit{Confidence amplifies explanation effects asymmetrically.} The benefit-to-harm ratio quantifies this misalignment starkly: 4.03 for overconfident novices (benefits dominate) versus 0.29 for humble experts (harms dominate)---a 14-fold difference that translates into divergent net welfare outcomes (+9.9 pp versus -0.72 pp). This pattern rewards unjustified confidence over appropriate caution, violating the principle that transparency should help those who need guidance while protecting those with sound judgment.

Table~\ref{tab:regression_accuracy} formalizes these patterns through mixed-effects regressions examining interaction effects.

\begin{table}[h!]
\centering \small
\caption{Interaction Analysis: Explanations, Confidence, and Competence}
\label{tab:regression_accuracy}
\begin{threeparttable}
\begin{tabular}{lccc}
\toprule
& \multicolumn{3}{c}{Posterior Diagnostic Accuracy (\%)} \\
\cmidrule(lr){2-4}
& AI Correct & AI Incorrect & Pooled \\
\midrule
\multicolumn{4}{l}{\textit{Main Effects}} \\
Explanation & 4.84** & -3.06 & 2.73** \\
& (1.95) & (2.05) & (1.31) \\
High Competence & 2.68 & 11.74** & 5.09*** \\
& (2.88) & (5.80) & (1.63) \\
High Confidence & -3.48 & -2.03 & -3.09 \\
& (4.05) & (3.77) & (2.94) \\
\midrule
\multicolumn{4}{l}{\textit{Two-way Interactions}} \\
Explanation $\times$ High Competence & -1.30 & -9.33 & -3.44* \\
& (3.36) & (6.25) & (1.96) \\
Explanation $\times$ High Confidence & 10.08** & -0.63 & 7.22** \\
& (4.40) & (4.42) & (3.17) \\
High Competence $\times$ High Confidence & 7.18 & -1.92 & 4.75 \\
& (4.94) & (7.11) & (3.34) \\
\midrule
\multicolumn{4}{l}{\textit{Three-way Interaction}} \\
Explanation $\times$ Comp. $\times$ Conf. & -8.69 & 7.14 & -4.47 \\
& (5.36) & (7.98) & (3.66) \\
\midrule
Constant (Low Comp, Low Conf, No Expl.) & 85.25*** & 10.63*** & 65.35*** \\
& (1.36) & (1.55) & (0.92) \\
\midrule
Individual Controls & \checkmark & \checkmark & \checkmark \\
Observations & 2,827 & 1,028 & 3,855 \\
\bottomrule
\end{tabular}
\begin{tablenotes}
\small
\item \textit{Notes:} Mixed-effects model with physician random effects. High Competence/Confidence indicate above-median values on prior accuracy and prior SSQ, respectively. Baseline group: low-competence, low-confidence physicians without explanations. Individual controls include gender, age, medical experience, and CRT score. Standard errors clustered at physician level. *** $p<0.01$, ** $p<0.05$, * $p<0.10$.
\end{tablenotes}
\end{threeparttable}
\end{table}

The regression analysis confirms the misalignment mechanistically. The Explanation $\times$ High Confidence interaction is large and positive when AI is correct (+10.08 pp, $p<0.05$), indicating confident physicians extract substantially more value from explanations regardless of their actual competence. Conversely, the Explanation $\times$ High Competence interaction is negative in the pooled model (-3.44 pp, $p<0.10$) and strongly negative when AI errs (-9.33 pp). Competent physicians who can identify incorrect AI advice without explanations (main effect: +11.74 pp when AI incorrect, $p<0.05$) lose this protective skepticism when persuasive explanations are provided. The three-way interaction, while not statistically significant, reinforces that being both competent and confident does not rescue physicians from explanation-induced harms.

This confidence-competence paradox reflects a clinical manifestation of the Dunning-Kruger effect \citep{kruger1999unskilled}. Overconfident novices---who lack metacognitive awareness of their limitations---benefit from explanations that correct their errors by leveraging their receptivity to authoritative-sounding guidance. Meanwhile, competent physicians with appropriate uncertainty---who possess the judgment to be appropriately skeptical---are misled by persuasive but incorrect explanations they would otherwise reject. Explanations thus amplify rather than mitigate the confidence-competence gap, creating perverse distributional effects where those who need help least benefit most.
\end{proof}

\bigskip
The confidence-competence paradox poses a fundamental challenge for transparency design. Targeting explanations to uncertain physicians would help uncertain novices but harm humble experts. Conversely, withholding explanations from confident physicians would protect calibrated experts but abandon overconfident novices who benefit most from corrective guidance. This suggests optimal transparency requires objective competence assessment rather than self-reported confidence---a challenging requirement that we address in the following policy analysis. Appendix Table~\ref{tab:combined_heterogeneity} provides detailed quartile analysis showing how this misalignment varies continuously across the skill distribution.

\subsection{Welfare Analysis and Policy Implications}
\label{sec:policy}

Given the explanation's dual effects depending on AI correctness, a natural question arises: what is the optimal AI transparency policy under this asymmetry? We leverage our experimental estimates to evaluate alternative transparency regimes through counterfactual welfare analysis, directly addressing whether AI systems should provide explanations universally (as mandated by the EU AI Act), selectively based on \emph{algorithmic confidence}, or conditionally on \emph{user competence}.

Consider a social planner choosing transparency policy $\tau \in \mathcal{T}$ to maximize diagnostic accuracy across a population of physician-patient interactions. The planner faces the fundamental trade-off identified in Result 1: explanations improve outcomes when AI is correct but harm them when incorrect. For any policy $\tau$, expected accuracy is:
\begin{equation}
\mathbb{E}[\text{Accuracy} \mid \tau] = \mu \cdot \Delta^+(\tau) + (1-\mu) \cdot \Delta^-(\tau) + \bar{A}
\label{eq:welfare}
\end{equation}
where $\mu = 0.73$ is AI accuracy in our setting, $\Delta^+(\tau)$ and $\Delta^-(\tau)$ are policy-specific treatment effects when AI is correct versus incorrect (measured from our experiment), and $\bar{A}$ is baseline accuracy without explanations. The welfare gain from policy $\tau$ relative to the status quo (no explanations) is:
\begin{equation}
W(\tau) = \mathbb{E}[\text{Accuracy} \mid \tau] - \bar{A} = \mu \cdot \Delta^+(\tau) + (1-\mu) \cdot \Delta^-(\tau)
\label{eq:welfare_gain}
\end{equation}

Table~\ref{tab:policy_welfare} presents welfare calculations for five alternative transparency policies. We translate accuracy improvements into economic value using conservative estimates of 500 million annual AI-assisted diagnoses globally and \$11,000 per diagnostic error \citep{tehrani2013diagnostic}.

\begin{table}[h!]
\centering \small
\caption{Welfare Effects of Alternative Transparency Policies}
\label{tab:policy_welfare}
\begin{threeparttable}
\begin{tabular}{lccccc}
\toprule
& \multicolumn{2}{c}{Treatment Effects} & \multicolumn{3}{c}{Welfare Outcomes} \\
\cmidrule(lr){2-3} \cmidrule(lr){4-6}
Policy & $\Delta^+$ & $\Delta^-$ & $W(\tau)$ & Economic & Efficiency \\
& (AI Correct) & (AI Incorrect) & (pp) & Value (\$B) & Ratio \\
\midrule
\textit{1. Status Quo} & 0 & 0 & 0 & 0 & --- \\
\quad (No explanations) & & & & & \\
\addlinespace
\textit{2. Universal Transparency} & +6.3 & -4.9 & +3.3 & +1.82 & 0.52 \\
\quad (EU AI Act mandate) & (0.8) & (1.6) & (1.3) & & \\
\addlinespace
\textit{3. Confidence Threshold} & +6.8 & -2.5 & +4.4 & +2.42 & 0.70 \\
\quad (Explain if AI conf. $>85\%$) & (0.9) & (1.9) & (1.4) & & \\
\addlinespace
\textit{4. Competence-Adaptive} & +7.6 & -3.4 & +4.7 & +2.59 & 0.75 \\
\quad (Low-comp only) & (2.7) & (3.6) & (2.1) & & \\
\addlinespace
\textit{5. First-Best} & +6.3 & 0 & +6.3 & +3.47 & 1.00 \\
\quad (Perfect foresight) & (0.8) & (0) & (0.8) & & \\
\bottomrule
\end{tabular}
\begin{tablenotes}
\small
\item \textit{Notes:} Treatment effects from experimental data (pooling deterministic and probabilistic formats). Standard errors clustered at the physician level in parentheses. Economic value = $W(\tau) \times 500\text{M diagnoses} \times \$11,000\text{ per error}$. Efficiency ratio = $W(\tau) / W(\text{First-Best})$ measures percentage of maximum achievable welfare. Policy 3: Explanations only when AI model confidence exceeds 85th percentile (conditional accuracy 74.4\%, applies to 52\% of cases). Policy 4: Explanations only for low-competence physicians (below-median baseline diagnostic accuracy, 50\% of sample), using effect estimates from overconfident and uncertain novices in Table~\ref{tab:confidence_competence_2x2}. Policy 5: Hypothetical theoretical benchmark providing explanations \& advice only when AI is correct (infeasible without perfect foresight). *** $p<0.01$, ** $p<0.05$, * $p<0.10$.
\end{tablenotes}
\end{threeparttable}
\end{table}

Three key findings emerge from this welfare analysis. First, universal transparency is inefficient. The EU AI Act's approach of mandatory explanations (Policy 2) achieves only 52\% of the first-best welfare gain. While generating \$1.82 billion in annual economic value, it leaves substantial welfare on the table (\$1.68 billion) due to symmetric application of explanations regardless of AI quality. The net welfare gain of +3.3 pp masks the fundamental tension: explanations help in 73\% of cases but harm in 27\%, with the aggregate benefit only weakly positive.

Second, contingent policies substantially improve welfare. Both confidence-based (Policy 3) and competence-based (Policy 4) contingent policies significantly outperform universal transparency. The confidence threshold policy achieves 70\% efficiency by concentrating explanations on high-confidence AI predictions (where accuracy is 74.4\%), generating \$2.42 billion in value---33\% more than universal transparency. Most strikingly, the competence-adaptive policy achieves 75\% efficiency (\$2.59 billion in value)---43\% higher than universal transparency---by targeting explanations to low-competence physicians who benefit from corrective guidance (overconfident and uncertain novices from Table~\ref{tab:confidence_competence_2x2}) while protecting high-competence physicians from misleading explanations.

Third, simple rules capture most gains. The competence-adaptive policy achieves 75\% efficiency (\$2.59B) and comes remarkably close to the first-best benchmark (\$3.5B), despite using only a binary competence classification based on baseline diagnostic accuracy. This suggests that even coarse targeting based on observable physician characteristics can dramatically improve welfare relative to universal transparency mandates. The small remaining gap to first-best (\$0.9B) represents the welfare loss from occasionally providing explanations when AI is incorrect within the low-competence group---a loss that could be further reduced by combining competence and confidence thresholds.

Our analysis suggests three concrete regulatory reforms that could generate an additional \$0.78 billion in annual welfare gains in healthcare alone (comparing competence-adaptive \$2.59B to universal transparency \$1.82B). First, move beyond universal mandates. The EU AI Act's blanket transparency requirement for high-risk systems ignores the fundamental trade-off we document: explanations help in 73\% of cases but systematically harm in 27\%. Regulators should permit \emph{conditional transparency based on algorithmic confidence thresholds}, providing explanations only when AI predictions exceed the 85th percentile of model certainty. Second, enable \emph{competence-adaptive systems}. Healthcare AI should adjust explanation provision based on user expertise, estimated from historical performance or certification levels. While this requires updating privacy regulations to allow performance tracking, the welfare gains justify these information costs---our estimates show competence-based targeting achieves 75\% efficiency versus 52\% for universal mandates. Third, require impact evaluation through randomized trials. Rather than mandating specific transparency mechanisms, regulators should require developers to demonstrate that explanations improve decision quality in their deployment context, given that transparency effects vary systematically with AI accuracy, advice format, and user characteristics.

These welfare estimates are conservative. External validity depends on how medical students generalize to practicing physicians, but time-pressed practitioners facing cognitive load likely exhibit greater susceptibility to explanation-induced over-reliance than our sample. This suggests the \$0.78 billion efficiency gain from contingent policies represents a lower bound. The critical policy insight is not that transparency is harmful per se, but that \emph{strategic silence}---withholding explanations when they mislead---is optimal. Current regulatory frameworks treat transparency as uniformly beneficial, mandating disclosure regardless of consequences. Our evidence demonstrates this assumption is wrong: the welfare case for contingent transparency is clear, and the challenge is crafting implementable rules that capture these gains while remaining politically acceptable to professional organizations and protective of patient welfare.

\section{Discussion and Conclusion}
\label{sec:conclusion}

This paper documents a fundamental trade-off in algorithmic transparency that challenges current regulatory approaches to AI governance. Using a lab-in-the-field experiment with 257 medical students making 3,855 incentivized diagnostic decisions, we establish the AI Transparency Paradox: explanations for AI recommendations improve outcomes when algorithms are correct but systematically worsen them when algorithms err. This asymmetry creates substantial welfare costs when algorithms are fallible---precisely the setting where transparency is most often mandated.

\subsection{Summary of Findings and Theoretical Contributions}

Our experimental evidence establishes five core results. First, explanations increase diagnostic accuracy by 6.3 percentage points when AI advice is correct but decrease accuracy by 4.9 percentage points when incorrect, creating an 11.2 percentage point paradox magnitude. Given realistic AI accuracy of 73\%, the net welfare effect is positive but modest (+3.3 percentage points), demonstrating that transparency's value depends critically on algorithmic quality---a dependence absent from current regulatory frameworks.

Second, conventional mitigation strategies fail. Combining explanations with probabilistic outputs amplifies rather than mitigates the harm of incorrect recommendations, with negative effects more than doubling in probabilistic formats (-6.6 pp vs. -3.1 pp). This suggests explanations resolve the ambiguity of uncertain signals into persuasive but potentially misleading guidance.

Third, we identify three complementary mechanisms. Explanations induce \textit{over-reliance}, causing physicians to update beliefs as if AI accuracy were 88.2\% when correct (15.2 pp above truth) and 79.2\% when incorrect (6.2 pp above truth). They impair \textit{discernment}, with physicians responding more to incorrect advice (+0.084 revision) than correct advice (+0.030). Finally, they create \textit{false confidence}, inflating certainty by 4.6 pp even when AI is wrong.

Fourth, the paradox creates perverse distributional effects. Explanations benefit overconfident novices most (+9.9 pp) while offering minimal gains to humble experts (+1.0 pp). When AI errs, humble experts suffer the largest harm (-12.4 pp) as explanations override their justified skepticism, rewarding unjustified confidence over sound judgment.

Fifth, contingent transparency policies substantially outperform universal mandates. Targeting explanations to low-competence physicians generates \$2.59 billion in annual value (75\% efficiency), saving \$0.78 billion annually in healthcare alone compared to universal transparency's \$1.82 billion.

Our findings extend information design theory \citep{kamenica2011bayesian} to settings where signal richness and signal value diverge. Unlike strategic senders who optimize information disclosure, AI systems mechanically generate explanations that symmetrically amplify both correct and incorrect advice through a persuasiveness function $\psi(e, x, \theta) = \exp(\lambda \cdot q(e))$. The critical insight is that while physicians treat explanation quality $q(e)$ as diagnostic of AI correctness, modern language models generate equally fluent explanations regardless of recommendation quality: $\mathbb{E}[q(e) | x = \theta] = \mathbb{E}[q(e) | x \neq \theta]$. This mismatch drives systematic over-reliance that persists even for incorrect advice, violating the standard premise that information has non-negative value.

\subsection{Policy Implications}

Our findings challenge the regulatory consensus embodied in the EU AI Act \citep{EUAIAct2024}, U.S. FDA guidelines, and emerging Chinese frameworks. Current regulations mandate explanations for high-risk AI systems under the assumption that understanding algorithmic reasoning universally improves collaboration. Our evidence demonstrates this assumption is incorrect: universal transparency imposes substantial welfare costs when algorithms are fallible.

We propose three regulatory reforms. First, move beyond universal mandates to permit conditional transparency based on algorithmic confidence thresholds. Requiring explanations only when AI predictions exceed the 85th percentile of model confidence (conditional accuracy 74.4\%) achieves 70\% efficiency while protecting users from misleading explanations. Second, enable competence-adaptive systems that adjust explanation provision based on user expertise, achieving 75\% efficiency by targeting explanations to low-competence users who benefit from correction while protecting high-competence users from misleading advice. Third, require impact evaluation through randomized trials rather than mandating specific transparency mechanisms, given that transparency effects vary systematically with AI accuracy, advice format, and user characteristics.

The rise of large language models intensifies the transparency paradox. Systems like GPT-4, 5 generate sophisticated explanations maintaining consistent fluency regardless of underlying uncertainty \citep{bubeck2023sparks}, \emph{decoupling explanation quality from recommendation reliability}. This amplifies the mechanisms we document: as AI capabilities advance, explanations become more persuasive for both correct and incorrect recommendations, \emph{strengthening rather than resolving the paradox}. Our theoretical framework, modeling explanations as persuasiveness weights that symmetrically amplify signals, reveals this extends beyond healthcare to any domain with fallible AI assistance, including criminal sentencing \citep{kleinberg2018human}, lending \citep{fuster2022predictably}, and hiring \citep{cowgill2020biased}. While transparency serves important accountability functions—enabling auditing, detecting bias, facilitating appeals—our evidence demonstrates these benefits must be weighed against decision-making harms. Current regulatory approaches sacrifice substantial welfare by prioritizing universal transparency when interpretability and welfare conflict. Contingent transparency frameworks offer a middle path: providing explanations when beneficial while protecting users from misleading guidance when algorithms err, achieving near-optimal welfare while maintaining accountability.

Our study has limitations suggesting future research directions. Our medical student sample may differ from experienced practitioners, though this likely biases estimates downward as time-pressed physicians may be more susceptible to over-reliance. Our controlled experimental setting enhances internal validity but requires replication in field settings. Our AI system has 73\% accuracy, realistic for current medical AI but below human expert performance. However, our theoretical framework predicts the paradox persists whenever AI is fallible: even 90\% accurate algorithms generate incorrect recommendations 10\% of the time. Finally, we examine one-shot decisions rather than repeated interactions enabling learning, though learning requires observing both recommendations and outcomes, often unavailable in medical practice where diagnostic uncertainty persists.

\subsection{Conclusion}

As regulators worldwide converge on mandating algorithmic transparency, our experimental evidence challenges this consensus by documenting the AI Transparency Paradox: explanations improve outcomes when algorithms are correct but systematically worsen them when algorithms err. This fundamental trade-off arises because modern language models generate equally convincing explanations regardless of recommendation quality, yet users treat explanation quality as diagnostic of accuracy—driving over-reliance, discernment failure, and false confidence that create substantial welfare costs. Our welfare analysis reveals that contingent transparency policies—selectively providing explanations based on AI confidence or user competence—generate 43\% more value than universal mandates, saving \$0.78 billion annually in healthcare alone (Table~\ref{tab:policy_welfare}). The challenge for policymakers is abandoning the false dichotomy between universal transparency and complete opacity in favor of evidence-based, contingent approaches that maximize welfare while maintaining accountability. As AI systems become more capable and widely deployed, optimal governance requires moving beyond one-size-fits-all rules toward sophisticated policies that account for the complex interplay between algorithmic quality, explanation design, and human judgment.

\footnotesize 
\bibliography{bib}

\newpage
\begin{appendices}

\setcounter{table}{0}
\setcounter{figure}{0}
\renewcommand{\thetable}{A\arabic{table}}
\renewcommand{\thefigure}{A\arabic{figure}}

\appendix
\section{Additional tables}

We first report the balance tests across schemes among demographic variables and responses from
the post-experimental questionnaires. Table~\ref{tab:balancecheck} reports the mean and $p-values$ from Fisher’s exact
tests, examining the equality of distributions across three schemes. Comparisons in all categories
are not statistically significant. The overall test of significance across all variables yields F = 0.82.
\begin{table}[H]
    \centering
    \caption{Randomization balance checks for the experiment}
    \begin{threeparttable}
    \begin{tabular}{lccccccc}
    \toprule
    \textbf{Individual characteristics} &\multicolumn{1}{c}{\textbf{D}}&\multicolumn{1}{c}{\textbf{\ED{}}}&\multicolumn{1}{c}{\textbf{\P{}}}&\multicolumn{1}{c}{\textbf{\EP{}}}&\multirow{1}*{\textbf{Total}}&\multirow{1}*{\textbf{$p$-value} }\\
    \midrule 
    \multicolumn{6}{l}{\textit{Medical ability and demographics:}} \\
    Ability  &3.750   &3.562 &3.600 &3.921  &3.708 & 0.621 \\
    Female(\%) & 63.10 &54.70 & 60.00 &54.70& 58.40&  0.649\\
    Age  & 25.03 & 24.64 & 25.06  & 24.81   & 24.88 & 0.369      \\  
    Medical length  &4.546 & 4.453 &4.738 &4.531   & 4.568  &0.972 \\
    Education &3.546   &3.453 &3.738 &3.531 &3.568 & 0.059\\\\
    \multicolumn{6}{l}{\textit{Economic preferences and trust:}} \\
    Altruism &2.780  & 2.793 &2.680 &2.719  &2.743 &0.293 \\
    Trust     &3.828 &3.984 &3.769 &3.703 & 3.821 & 0.068  \\
    CRT &2.109  &1.922 &2.138 &2.062 &2.058 & 0.248\\\\
     \multicolumn{6}{l}{\textit{Algorithm preference surveys:}} \\
    AI Trust  &3.934 & 3.940  &3.815 &3.734  &3.856 & 0.760\\ 
    Fairness  & 3.275 & 3.403  &3.301 &3.275  &3.314 &0.561 \\
    Literacy &3.581 &3.556 &3.360 &3.597  &3.522 & 0.596\\
    Awareness &4.184 &4.134 &3.969 &4.037  &4.081& 0.362\\\\
   
    Observations & 64&64&65&64&257&-\\
    \bottomrule
    \end{tabular}
    % }
    \begin{tablenotes}
        \small
        \item \textit{Notes:}  This table reports the average values of individual characteristic variables across treatments. $p-$values are from the Fisher's exact test, treating each individual as an independent observation. There were no significant differences in the distribution of individual characteristics between treatments. For an overall test of balance across the 12 variables, F-statistic=0.82 ($p$=0.62).
        \end{tablenotes}
    \end{threeparttable}
    \label{tab:balancecheck}
    \end{table}

\newpage
Table~\ref{tab:main_paradox_appendix} and \ref{tab:format_robustness_appendix}  presents the paradox conditional on whether the AI advice was deterministic or probabilistic and they are robust to mixed-effects regressions with physician random effects.

\begin{table}[H]
\centering
\caption{The AI Transparency Paradox: Diagnostic Accuracy by AI Correctness and Explanation Status}
\label{tab:main_paradox_appendix}
\begin{threeparttable}
\begin{tabular}{lcccc}
\toprule
& \multicolumn{2}{c}{\textbf{AI Advice is Correct}} & \multicolumn{2}{c}{\textbf{AI Advice is Incorrect}} \\
\cmidrule(lr){2-3} \cmidrule(lr){4-5}
& No Explanation & Explanation & No Explanation & Explanation \\
\midrule
\textbf{Diagnostic Accuracy} & 87.4\% & 93.7\% & 14.3\% & 9.4\% \\
\textbf{Standard Error} & (0.6) & (0.4) & (1.2) & (1.0) \\
\textbf{Observations} & 1,419 & 1,408 & 516 & 512 \\
\midrule
\textbf{Explanation Effect} & \multicolumn{2}{c}{+6.0 pp***} & \multicolumn{2}{c}{-4.4 pp**} \\
\textbf{95\% CI} & \multicolumn{2}{c}{[3.6, 8.5]} & \multicolumn{2}{c}{[-8.6, -0.3]} \\
\midrule
\textbf{Overall Effect} & \multicolumn{4}{c}{+3.2 pp***} \\
\textbf{Net Benefit} & \multicolumn{4}{c}{73\% × 6.0pp - 27\% × 4.4pp = +3.2 pp} \\
\bottomrule
\end{tabular}
\begin{tablenotes}
\small
\item \textit{Notes:} 
Diagnostic accuracy is the probability assigned to the correct diagnosis. Standard errors in parentheses, clustered at the physician level. Explanation effects demonstrate the results of mixed-effects regression analysis comparing the explanatory and non-explanatory conditions within each AI correctness category.  *** p<0.01, ** p<0.05, * p<0.10.
\end{tablenotes}
\end{threeparttable}
\end{table}

\begin{table}[H]
\centering
\caption{Transparency Paradox Robustness Across AI Advice Formats: Diagnostic Accuracy by AI Correctness, Explanation Status and Physicians' prior Correctness}
\label{tab:format_robustness_appendix}
\begin{threeparttable}
\setlength{\tabcolsep}{3mm}{
\begin{tabular}{lcccc}
\toprule
& \multicolumn{2}{c}{\textbf{Deterministic Format}} & \multicolumn{2}{c}{\textbf{Probabilistic Format}} \\
\cmidrule(lr){2-3} \cmidrule(lr){4-5}
& AI Correct & AI Incorrect & AI Correct & AI Incorrect \\
\midrule
\textbf{No Explanation} & 89.0\% & 10.4\% & 85.7\% & 18.2\% \\
& (0.9) & (1.5) & (0.9) & (1.9) \\
\textbf{With Explanation} & 95.6\% & 7.3\% & 91.7\% & 11.6\% \\
& (0.6) & (1.4) & (0.5) & (1.4) \\
\midrule
\textbf{{Explanation Effect}} & +6.6*** & -2.9 & +4.2*** & -7.2*** \\
& (1.2) & (2.4) & (1.6) & (3.3) \\
\midrule
\textbf{Paradox Magnitude} & \multicolumn{2}{c}{9.7 pp} & \multicolumn{2}{c}{12.6 pp} \\
\textbf{Net Benefit} & \multicolumn{2}{c}{3.9pp***} & \multicolumn{2}{c}{1.2pp} \\
\midrule
\textbf{Observations} & 1,408 & 512 & 1,419 & 516 \\
\bottomrule
\end{tabular}}
\begin{tablenotes}
\small
\item \textit{Notes:} Deterministic format presents single recommendation (``AI recommends diagnosis B''). Probabilistic format presents probability distribution (``60\% B, 30\% A, 10\% C''). Explanation effects demonstrate the results of mixed-effects regression analysis. *** p<0.01, ** p<0.05, * p<0.10.
\end{tablenotes}
\end{threeparttable}
\end{table}

\newpage
Table \ref{tab:combined_heterogeneity} examines the heterogeneity of the transparency paradox along two dimensions: prior confidence and prior competence. Explanations substantially increase physicians’ accuracy when the AI is correct, but reduce their accuracy when the AI is incorrect. Among high-confidence physicians, however, explanations produce no noticeable difference when the AI is incorrect.

\begin{table}[H]
\centering \small
\caption{Heterogeneity: The Paradox by Prior Confidence and Prior Competence}
\label{tab:combined_heterogeneity}
\begin{threeparttable}
\begin{tabular}{lccccc}
\toprule
& \multicolumn{4}{c}{\textbf{Quartile}} \\
\cmidrule(lr){2-5}
& Q1 (Low) & Q2 & Q3 & Q4 (High) \\
\midrule
\multicolumn{5}{l}{\textbf{Panel A: Classification by Prior Confidence (SSQ)}} \\
\multicolumn{5}{l}{\textit{When AI is Correct:}} \\
No Explanation & 83.9\% & 87.6\% & 88.9\% & 89.3\% \\
& (1.1) & (1.2) & (1.3) & (1.5) \\
With Explanation & 89.0\% & 92.2\% & 96.6\% & 96.6\% \\
& (0.8) & (0.9) & (0.5) & (0.8) \\
\textbf{Effect} & +5.1pp** & +4.6pp** & +7.6pp*** & +7.3pp*** \\
\multicolumn{5}{l}{\textit{When AI is Incorrect:}} \\
No Explanation & 11.0\% & 14.8\% & 17.1\% & 14.6\% \\
& (1.6) & (2.4) & (2.8) & (2.9) \\
With Explanation & 7.3\% & 9.5\% & 9.3\% & 11.6\% \\
& (1.1) & (1.9) & (2.1) & (2.6) \\
\textbf{Effect} & -3.7pp** & -5.3pp** & -7.8pp** & -3.0pp \\
\textbf{Paradox Magnitude} & 8.8pp & 9.9pp & 15.4pp & 10.3pp \\
\midrule
\multicolumn{5}{l}{\textbf{Panel B: Classification by Prior Accuracy (Competence)}} \\
\multicolumn{5}{l}{\textit{When AI is Correct:}} \\
No Explanation & 83.8\% & 85.3\% & 90.1\% & 91.5\% \\
& (1.2) & (1.5) & (1.2) & (1.1) \\
With Explanation & 89.0\% & 94.3\% & 94.3\% & 95.7\% \\
& (1.1) & (0.7) & (0.7) & (0.7) \\
\textbf{Effect} & +5.2pp** & +9.0pp*** & +4.2pp*** & +4.2pp*** \\
\multicolumn{5}{l}{\textit{When AI is Incorrect:}} \\
No Explanation & 9.2\% & 11.5\% & 13.3\% & 25.5\% \\
& (0.8) & (1.3) & (2.4) & (4.4) \\
With Explanation & 6.6\% & 6.9\% & 8.1\% & 15.5\% \\
& (1.4) & (1.5) & (1.7) & (2.8) \\
\textbf{Effect} & -2.6pp* & -4.6pp** & -5.2pp* & -10.0pp** \\
\textbf{Paradox Magnitude} & 7.8pp & 13.6pp & 9.4pp & 14.2pp \\
\midrule
\multicolumn{5}{l}{\textbf{Panel C: Key Distinctions}} \\
\textbf{Benefit/Harm Ratio:} & & & & \\
\quad By Confidence (SSQ) & 1.38 & 0.87 & 0.97 & 2.43 \\
\quad By Competence (Accuracy) & 2.00 & 1.96 & 0.81 & 0.42 \\
\textbf{Net Welfare Effect:} & & & & \\
\quad By Confidence & +2.6pp & +1.5pp & +3.4pp & +4.5pp \\
\quad By Competence & +2.7pp & +5.1pp & +0.7pp & -2.0pp \\
\midrule
Observations & 975 & 960 & 975 & 945 \\
\bottomrule
\end{tabular}
\begin{tablenotes}
\small
\item \textit{Notes:} Panel A classifies physicians by prior confidence (SSQ of initial beliefs). Panel B classifies by prior competence (actual diagnostic accuracy). Paradox magnitude = |help| + |harm|. Benefit/harm ratio = help effect / |harm effect|. Net welfare calculated using 73\% AI accuracy rate. Standard errors in parentheses. *** p<0.01, ** p<0.05, * p<0.10.
\end{tablenotes}
\end{threeparttable}
\end{table}

\newpage

\section{Instructions to Participants}
\label{sec:append_instructions}

\subsection*{Welcome}
Welcome to this decision-making experiment. Please make your decisions carefully, as you will be paid based on the choices you make. 
Additionally, based on your decisions, a donation will be made to a patient-regarding charity.

\subsection*{The Scenario}
You will be presented with \textbf{15 medical scenarios} and asked to perform \textbf{diagnostic assessments} based on a patient’s condition. Each scenario includes \textbf{five possible diagnoses} (A, B, C, D, E), with \textbf{only one} correct. You will assign probabilities to each option, indicating how likely you believe each is correct.

\subsubsection*{\underline{The AI Advice}}
You will receive \textbf{AI advice} (generated by ChatGPT 4, accuracy $>70\%$) providing \textbf{[one/two] suggested option(s)} [and a short explanation]. The AI’s suggestion is not guaranteed to be correct.

\subsubsection*{\underline{Overview of Your Tasks}}
In each scenario, you will be instructed to complete \textbf{three tasks}:
\begin{itemize}
  \item \textbf{Task 1: Initial Estimations} (before seeing AI advice)  
  \item \textbf{Task 2: Expected Informativeness Score} (still before AI advice)  
  \item \textbf{Task 3: Final Estimations} (after seeing AI advice)
\end{itemize}
We use a standard quadratic scoring rule to determine your payoffs in each task (you may ask us the details after the experiment), this rule makes sure that being honest and accurate is in your best interest.

\subsection*{Task 1: Initial Estimations (Before AI Advice)}
\textbf{Your Task:}  
Read the scenario describing the patient’s condition. Distribute probabilities across the five options (A, B, C, D, E) indicating how likely you believe each is, all probabilities shall sum to 100\%.

\noindent\textbf{Payment:}  
At the end of the scenario, the correct diagnosis will be revealed. Your payoff depends on how close your stated probabilities are to the truth, evaluated by a \emph{quadratic scoring rule}. The maximum payment for this task is \(\Payment{}\) Yuan.

\begin{center}
  \includegraphics[width=0.3\textwidth]{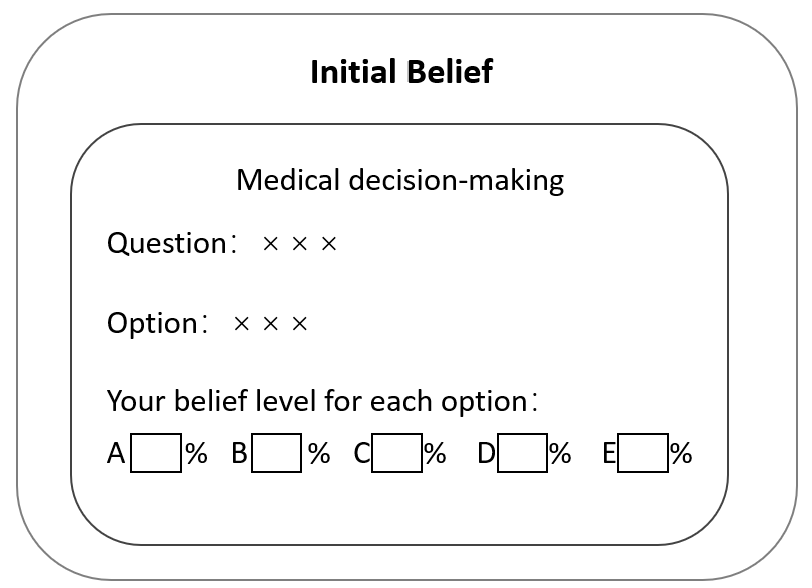}\\
  {\small Figure 1: Example Screenshot of Initial Estimation}
\end{center}

\subsection*{Task 2: Expected Informativeness Score (Before AI Advice)}
\textbf{Definition:}  
Your \emph{Informativeness Score} measures how certain (or “informative”) your probability estimates are. It is calculated as:

\[
  \sum_{t \in \{A,B,C,D,E\}} (\text{Initial Estimation}_t)^2.
\]
An even split of 20\% each yields $0.2^2\times5=0.2$, while being 100\% sure of one option yields $1^2=1$.
Hence, more confident assessments produce higher informativeness scores.

\noindent \textbf{Your Task:}  
Right after giving your initial estimates, you will see your informativeness score. Then, predict how you think \emph{this score will change} after viewing the AI advice.

\noindent \textbf{Payment:}  
You will be rewarded based on how accurate your prediction is. The maximum payment for this is \(\Payment{}\) Yuan.

\begin{center}
  \includegraphics[width=0.5\textwidth]{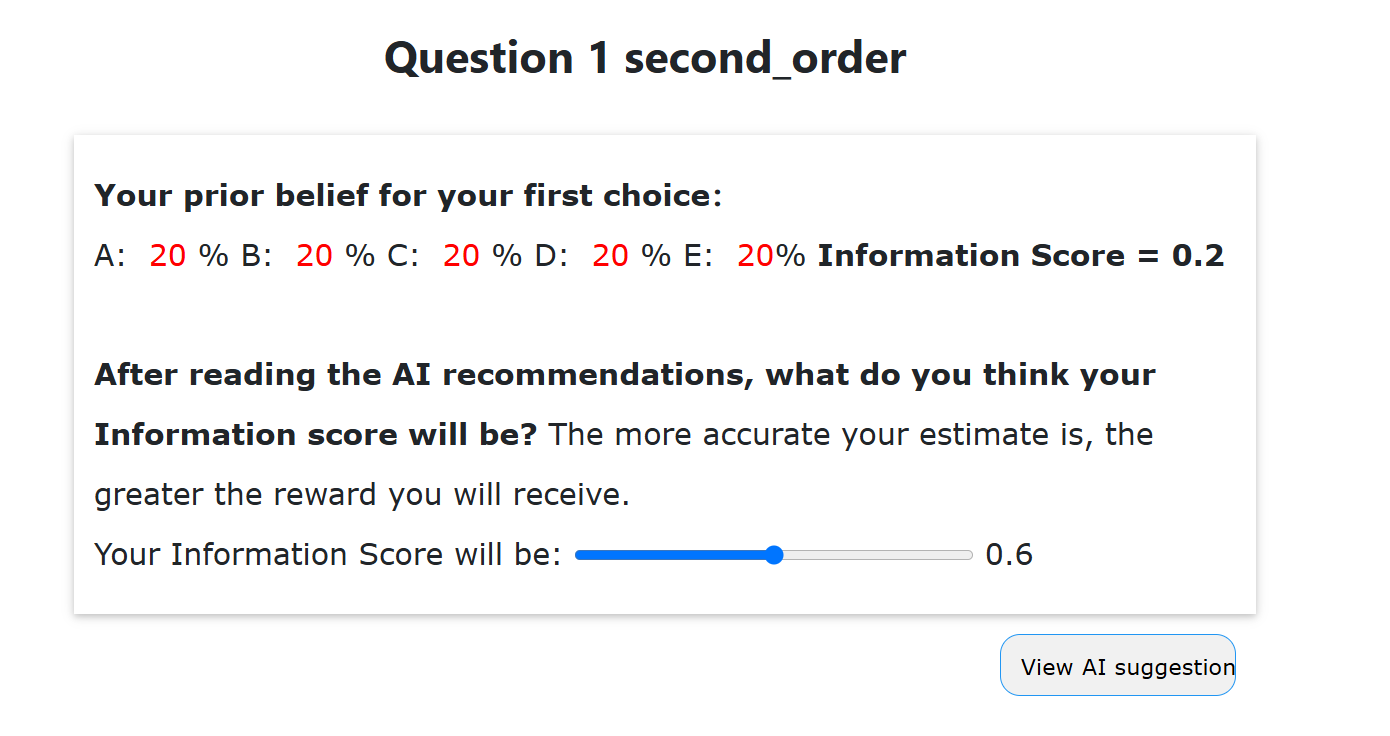}\\
  {\small Figure 2: Example Screenshot of Expected Informativeness Score}
\end{center}

\subsection*{Task 3: Final Estimations (After AI Advice)}
\textbf{Your Task:}  
After reading the AI’s [one/two] suggested option(s) [and explanation], assign probabilities again to the five options (A, B, C, D, E) so that they sum to 100\%.

\noindent \textbf{Payment:}  
Your final estimations is also scored using the quadratic scoring rule to make sure it is of your best interest to provide us your actural estimations. The maximum payment for this part is \(\Payment{}\) Yuan.

\begin{center}
  \includegraphics[width=0.3\textwidth]{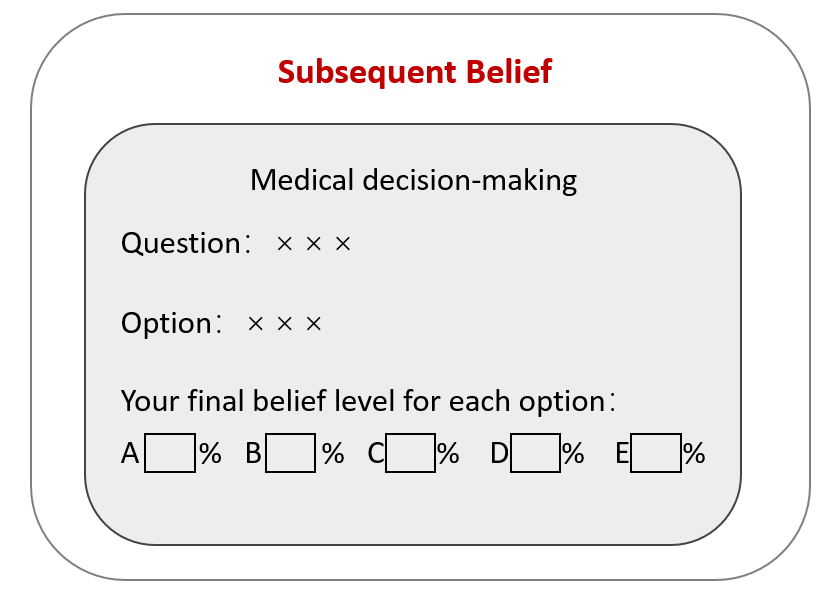}\\
  {\small Figure 3: Example Screenshot of Final Estimation}
\end{center}

\subsection*{Your Total Payment}
\textbf{Additional Payment:}  
At the start, you will answer \textbf{10 multiple-choice questions} (one correct option among five). Each correct answer earns you \textbf{1 Yuan}.

\noindent \textbf{Overall Payment:}  
You will receive payoffs for each of the 15 scenarios from:  
- Task 1 (Initial Estimations)  
- Task 2 (Expected Informativeness Score)  
- Task 3 (Final Estimations)  
plus any additional payments from the 10 initial questions.

\subsection*{Donation to Charity}
A donation will be made on your behalf to a patient-focused charity based on your payoffs in the Final Estimation Tasks. Higher earnings lead to a larger donation to help patients with ALS, up to a maximum of 10 Yuan.

Before the experiment begins, you will answer several questions to ensure you understand these instructions. Your responses \textbf{will not} affect your payment. If you have any questions, please raise your hand.

\begin{center}
    \textbf{Thank you for your participation and good luck!}   
\end{center}

\section*{Post-experimental questionnaire}

\noindent{\textbf{ Questionnaire 1}}

1. What is your gender?

\quad\scriptsize{\textcircled{}}\normalsize{} Male

\quad\scriptsize{\textcircled{}}\normalsize{} Female

2. What is your age?

\quad\scriptsize{\textcircled{}}\normalsize{} 18-20 years old

\quad\scriptsize{\textcircled{}}\normalsize{} 21-23 years old

\quad\scriptsize{\textcircled{}}\normalsize{} 24-26 years old

\quad\scriptsize{\textcircled{}}\normalsize{} 27-29 years old

\quad\scriptsize{\textcircled{}}\normalsize{} Over 30 years old

3. What is the duration of your education?

\quad\scriptsize{\textcircled{}}\normalsize{} 2 years or less

\quad\scriptsize{\textcircled{}}\normalsize{} 3 years

\quad\scriptsize{\textcircled{}}\normalsize{} 4 years

\quad\scriptsize{\textcircled{}}\normalsize{} 5 years

\quad\scriptsize{\textcircled{}}\normalsize{} More than 5 years

4. What is your highest education level?

\quad\scriptsize{\textcircled{}}\normalsize{} Secondary Specialized

\quad\scriptsize{\textcircled{}}\normalsize{} Associate Degree

\quad\scriptsize{\textcircled{}}\normalsize{} Bachelor's Degree

\quad\scriptsize{\textcircled{}}\normalsize{} Master's Degree

\quad\scriptsize{\textcircled{}}\normalsize{} doctorate

5. If you unexpectedly received 1000 yuan today, how much would you donate to charitable causes?

6. How willing are you to donate to public welfare causes without expecting anything in return?

\quad\scriptsize{\textcircled{}}\normalsize{} Extremely unlikely

\quad\scriptsize{\textcircled{}}\normalsize{} Unlikely

\quad\scriptsize{\textcircled{}}\normalsize{} Neutral

\quad\scriptsize{\textcircled{}}\normalsize{} Likely

\quad\scriptsize{\textcircled{}}\normalsize{} Extremely likely

7. I believe people are generally well-intentioned.

\quad\scriptsize{\textcircled{}}\normalsize{} Extremely unlikely

\quad\scriptsize{\textcircled{}}\normalsize{} Unlikely

\quad\scriptsize{\textcircled{}}\normalsize{} Neutral

\quad\scriptsize{\textcircled{}}\normalsize{} Likely

\quad\scriptsize{\textcircled{}}\normalsize{} Extremely likely

\vspace{5mm}
\noindent{\textbf{Questionnaire 2: Raven test}}
\vspace{3mm}

This test task contains 6 figures as shown below example, each figure consists of 3x3 different patterns,where the last pattern is blank. Each row/column of these patterns is arranged according to certain rules. Your task is to find the best pattern to fill in the blanks. You have a total of 5 minutes to complete this part of the test. For every correct answer, you will get an extra 1 yuan.

\begin{figure}[h!]
    \centering
    \includegraphics[width=0.4\textwidth]{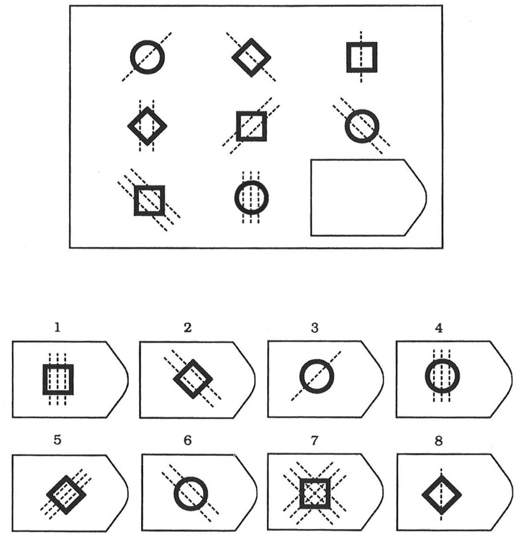}
   
    \caption{An example of a Raven task test}
    %\label{fig:bar}
\end{figure}

\vspace{5mm}
\noindent{\textbf{Questionnaire 3: CRT test}}
\vspace{3mm}

Please answer the following four questions. You have a total of 3 minutes to complete this part of the
test. You will receive an additional 1 yuan bonus for each correct answer of these questions.

1. A pair of tennis rackets and a ball together cost 1.10, and the racket is 1 more expensive than the ball. What is the price of the ball in dollars?

2. If 5 machines can produce 5 parts in 5 minutes, how many minutes would it take for 100 machines to produce 100 parts?

3. A barrel of pure water would be finished by Xiao Ming in 6 days and by Xiao Hong in 12 days. If Xiao Ming and Xiao Hong become roommates and drink from the same barrel, how many days would it take for them to finish the water?

4. As shown in the figure below (which is not provided here), there are four cards (A, B, C, D) on the table. Each card has a number on the front and a color on the back. Now, Xiao Ming has made the following conjecture: If the front of a card is an even number, then its back is blue. Assuming you can look at these cards, which cards must you turn over to verify whether Xiao Ming's conjecture is correct?
\begin{figure}[h!]
    \centering
    \includegraphics[width=0.6\textwidth]{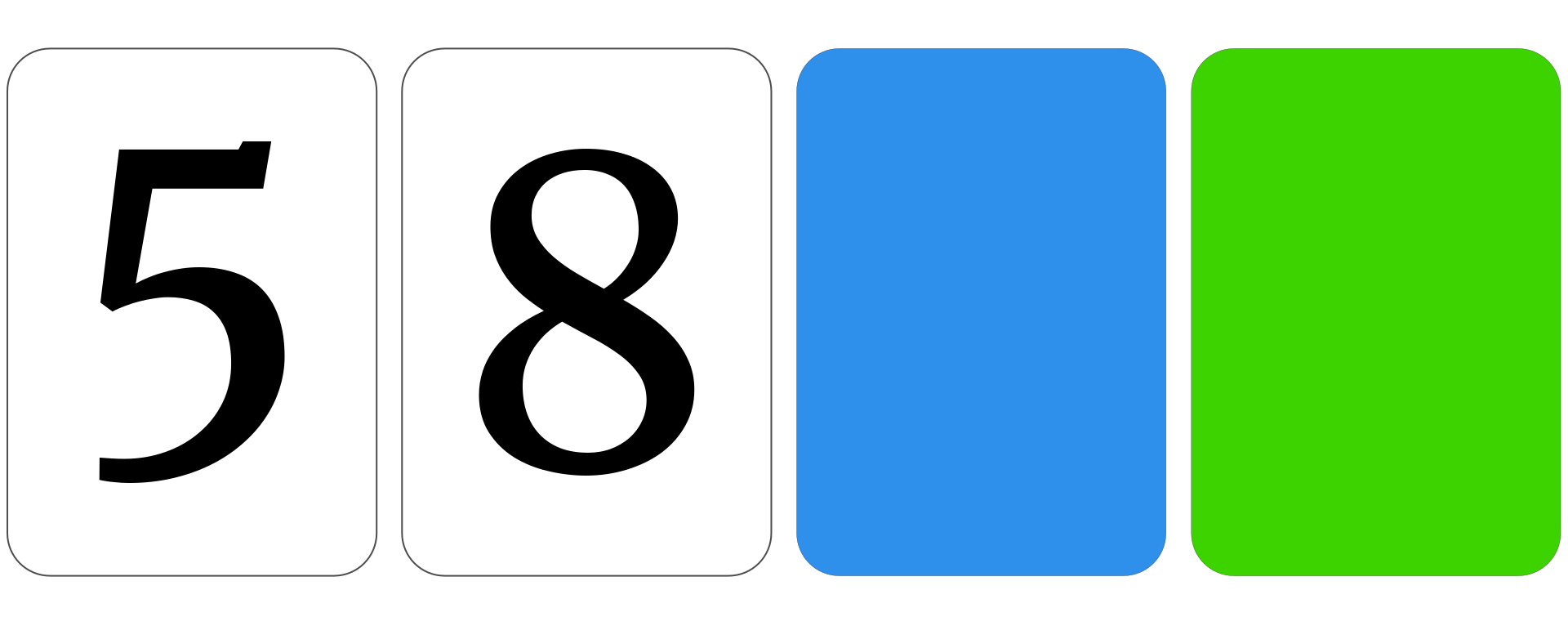}
    \caption{An example of a CRT test}
\end{figure}

\vspace{5mm}
\noindent{\textbf{Questionnaire 4: Algorithm literacy and awareness}}
\vspace{3mm}

Below are some descriptions related to the field of artificial intelligence. For each statement, please select according to your thoughts.

1. Are you aware of the current state of development in artificial intelligence (AI)?

\scriptsize{\textcircled{}}\normalsize{}  Completely Unaware \scriptsize{\textcircled{}}\normalsize{} Not Very Aware \scriptsize{\textcircled{}}\normalsize{} Somewhat Aware \scriptsize{\textcircled{}}\normalsize{} Aware \scriptsize{\textcircled{}}\normalsize{} Very Aware

2. Are you aware of the potential use of AI algorithms in medical decision-making?

\scriptsize{\textcircled{}}\normalsize{}  Completely Unaware \scriptsize{\textcircled{}}\normalsize{} Not Very Aware \scriptsize{\textcircled{}}\normalsize{} Somewhat Aware \scriptsize{\textcircled{}}\normalsize{} Aware \scriptsize{\textcircled{}}\normalsize{} Very Aware

3. Do you think AI algorithms will have what kind of impact on the future development of the medical industry?

\scriptsize{\textcircled{}}\normalsize{} Very Negative Impact
\scriptsize{\textcircled{}}\normalsize{} Negative Impact
\scriptsize{\textcircled{}}\normalsize{} No Impact
\scriptsize{\textcircled{}}\normalsize{} Positive Impact
\scriptsize{\textcircled{}}\normalsize{} Very Positive Impact

4. Are you aware of the basic principles of machine learning?

\scriptsize{\textcircled{}}\normalsize{}  Completely Unaware \scriptsize{\textcircled{}}\normalsize{} Not Very Aware \scriptsize{\textcircled{}}\normalsize{} Somewhat Aware \scriptsize{\textcircled{}}\normalsize{} Aware \scriptsize{\textcircled{}}\normalsize{} Very Aware

5. Do you believe that a critical thinking approach should be maintained when using AI algorithms?

\scriptsize{\textcircled{}}\normalsize{} Strongly Disagree
\scriptsize{\textcircled{}}\normalsize{} Disagree
\scriptsize{\textcircled{}}\normalsize{} Neutral
\scriptsize{\textcircled{}}\normalsize{} Agree
\scriptsize{\textcircled{}}\normalsize{} Strongly Agree

\vspace{5mm}
Below are some relevant descriptions of AI for the healthcare field, for each of the following statements, please choose as you see fit.

1. How useful do you think AI algorithms are in medical decision making?

\scriptsize{\textcircled{}}\normalsize{} Almost no effect
\scriptsize{\textcircled{}}\normalsize{} Limited role
\scriptsize{\textcircled{}}\normalsize{} Somewhat useful
\scriptsize{\textcircled{}}\normalsize{} Very useful

2. How useful do you think AI algorithms are in assisting physicians' treatment systems?

\scriptsize{\textcircled{}}\normalsize{} Almost no effect
\scriptsize{\textcircled{}}\normalsize{} Limited role
\scriptsize{\textcircled{}}\normalsize{} Somewhat useful
\scriptsize{\textcircled{}}\normalsize{} Very useful

3. How useful do you think AI algorithms are in healthcare data analysis?

\scriptsize{\textcircled{}}\normalsize{} Almost no effect
\scriptsize{\textcircled{}}\normalsize{} Limited role
\scriptsize{\textcircled{}}\normalsize{} Somewhat useful
\scriptsize{\textcircled{}}\normalsize{} Very useful

4. Have you ever questioned or validated the results of AI algorithms in healthcare decision-making?

\scriptsize{\textcircled{}}\normalsize{} Not at all
\scriptsize{\textcircled{}}\normalsize{} Only once
\scriptsize{\textcircled{}}\normalsize{} A few times
\scriptsize{\textcircled{}}\normalsize{} Regularly

5. Would you be willing to receive specialized training on the use of AI in healthcare?

\scriptsize{\textcircled{}}\normalsize{} Totally unwilling
\scriptsize{\textcircled{}}\normalsize{} Not very willing
\scriptsize{\textcircled{}}\normalsize{} Willing 
\scriptsize{\textcircled{}}\normalsize{} Very willing

\vspace{5mm}
\noindent{\textbf{Questionnaire 5: Algorithm fairness and trust}}
\vspace{3mm}

1. Which AI large language model have you used?

\quad $\Square$  ChatGPT

\quad $\Square$ Gemini

\quad $\Square$ Wen Xin Yi Yan

\quad $\Square$ Kimi chat

\quad $\Square$ Sora

\quad $\Square$ Dall.E3

\quad $\Square$ Other

\quad $\Square$ Don't know

If you selected ``Other", please enter your answer below.

2. How often do you use the Large Language Model?

\scriptsize{\textcircled{}}\normalsize{} Every day
\scriptsize{\textcircled{}}\normalsize{} 3 times a week
\scriptsize{\textcircled{}}\normalsize{} Once every six months
\scriptsize{\textcircled{}}\normalsize{} Never

3. Below are two different perspectives on Al for healthcare, for each of the following statements, please choose as you see fit. Whichever side of the argument you agree with, select the scale point that is close to that argument.(There are five points on the scale, and the meanings from left to right are ``Strongly Agree With Viewpoint A", ``Somewhat Agree With Viewpoint A", ``Neutral", ``Strongly Agree with Viewpoint B", ``Strongly Agree with Viewpoint B").

(1)View A : AI technology has made it more difficult for patients in low-income or remote areas to access quality healthcare.

\ \quad View B : AI technology makes it more likely that patients in low-income or remote areas will have access to high-quality healthcare.

\qquad Strongly Agree With  A  \scriptsize{\textcircled{}}\normalsize{}----\scriptsize{\textcircled{}}\normalsize{}----\scriptsize{\textcircled{}}\normalsize{}----\scriptsize{\textcircled{}}\normalsize{}----\scriptsize{\textcircled{}}\normalsize{}  Strongly Agree With  B

(2)View A : AI technology has a significant impact on healthcare access equity.

\ \quad View B : AI technology did not have any significant impact on equity of access to healthcare.

\qquad Strongly Agree With  A  \scriptsize{\textcircled{}}\normalsize{}----\scriptsize{\textcircled{}}\normalsize{}----\scriptsize{\textcircled{}}\normalsize{}----\scriptsize{\textcircled{}}\normalsize{}----\scriptsize{\textcircled{}}\normalsize{}  Strongly Agree With  B

(3)View A : AI technology makes resources tend to be0allocated to patients or providers who can pay higher fees.

\ \quad View B : AI technology has led to a more even distribution of resources.

\qquad Strongly Agree With  A  \scriptsize{\textcircled{}}\normalsize{}----\scriptsize{\textcircled{}}\normalsize{}----\scriptsize{\textcircled{}}\normalsize{}----\scriptsize{\textcircled{}}\normalsize{}----\scriptsize{\textcircled{}}\normalsize{}  Strongly Agree With  B

(4)View A : The use of AI technology may exacerbate inequalities in research and development of effective treatments for certain diseases.

\ \quad View B : AI technology has made it possible for rare diseases to be more fully researched as well.

\qquad Strongly Agree With  A  \scriptsize{\textcircled{}}\normalsize{}----\scriptsize{\textcircled{}}\normalsize{}----\scriptsize{\textcircled{}}\normalsize{}----\scriptsize{\textcircled{}}\normalsize{}----\scriptsize{\textcircled{}}\normalsize{}  Strongly Agree With  B

(5)View A : AI technology may cause physicians to become overly reliant on technology at the expense of direct patient interaction.

\ \quad View B : AI technology may increase patient trust in healthcare because it provides more accurate medical information.

\qquad Strongly Agree With  A  \scriptsize{\textcircled{}}\normalsize{}----\scriptsize{\textcircled{}}\normalsize{}----\scriptsize{\textcircled{}}\normalsize{}----\scriptsize{\textcircled{}}\normalsize{}----\scriptsize{\textcircled{}}\normalsize{}  Strongly Agree With  B

(6)View A : AI provides inaccurate results.

\ \quad View B : AI provides accurate results.

\qquad Strongly Agree With  A  \scriptsize{\textcircled{}}\normalsize{}----\scriptsize{\textcircled{}}\normalsize{}----\scriptsize{\textcircled{}}\normalsize{}----\scriptsize{\textcircled{}}\normalsize{}----\scriptsize{\textcircled{}}\normalsize{}  Strongly Agree With  B

(7)View A : AI provides results that are not easily applied to common problems.

\ \quad View B : AI provides enough results to apply to common problems.

\qquad Strongly Agree With  A  \scriptsize{\textcircled{}}\normalsize{}----\scriptsize{\textcircled{}}\normalsize{}----\scriptsize{\textcircled{}}\normalsize{}----\scriptsize{\textcircled{}}\normalsize{}----\scriptsize{\textcircled{}}\normalsize{}  Strongly Agree With  B

(8)View A : If hospitals introduce AI-assisted medical technology, I'm not so sure about using AI for assisted diagnosis and treatment.

\ \quad View B : If the hospital introduces AI-assisted healthcare technology, I will combine it with a large model to assist in the consultation.

\qquad Strongly Agree With  A  \scriptsize{\textcircled{}}\normalsize{}----\scriptsize{\textcircled{}}\normalsize{}----\scriptsize{\textcircled{}}\normalsize{}----\scriptsize{\textcircled{}}\normalsize{}----\scriptsize{\textcircled{}}\normalsize{}  Strongly Agree With  B

(9)View A : Using the Large Language Model will not improve the accuracy of a physician's diagnosis.

\ \quad View B : Using the Large Language Model can improve the accuracy of a physician's diagnosis.

\qquad Strongly Agree With  A  \scriptsize{\textcircled{}}\normalsize{}----\scriptsize{\textcircled{}}\normalsize{}----\scriptsize{\textcircled{}}\normalsize{}----\scriptsize{\textcircled{}}\normalsize{}----\scriptsize{\textcircled{}}\normalsize{}  Strongly Agree With  B

(10)View A : AI algorithms have a very negative impact on the future of the healthcare industry.

\quad\ \, View B : AI algorithms have a very positive impact on the future of the healthcare industry.

\qquad Strongly Agree With  A  \scriptsize{\textcircled{}}\normalsize{}----\scriptsize{\textcircled{}}\normalsize{}----\scriptsize{\textcircled{}}\normalsize{}----\scriptsize{\textcircled{}}\normalsize{}----\scriptsize{\textcircled{}}\normalsize{}  Strongly Agree With  B

\clearpage
\section{Donation Certificate}
\label{sec:donation}
\begin{figure}[h!]
    \centering
    \includegraphics[width=0.6\linewidth]{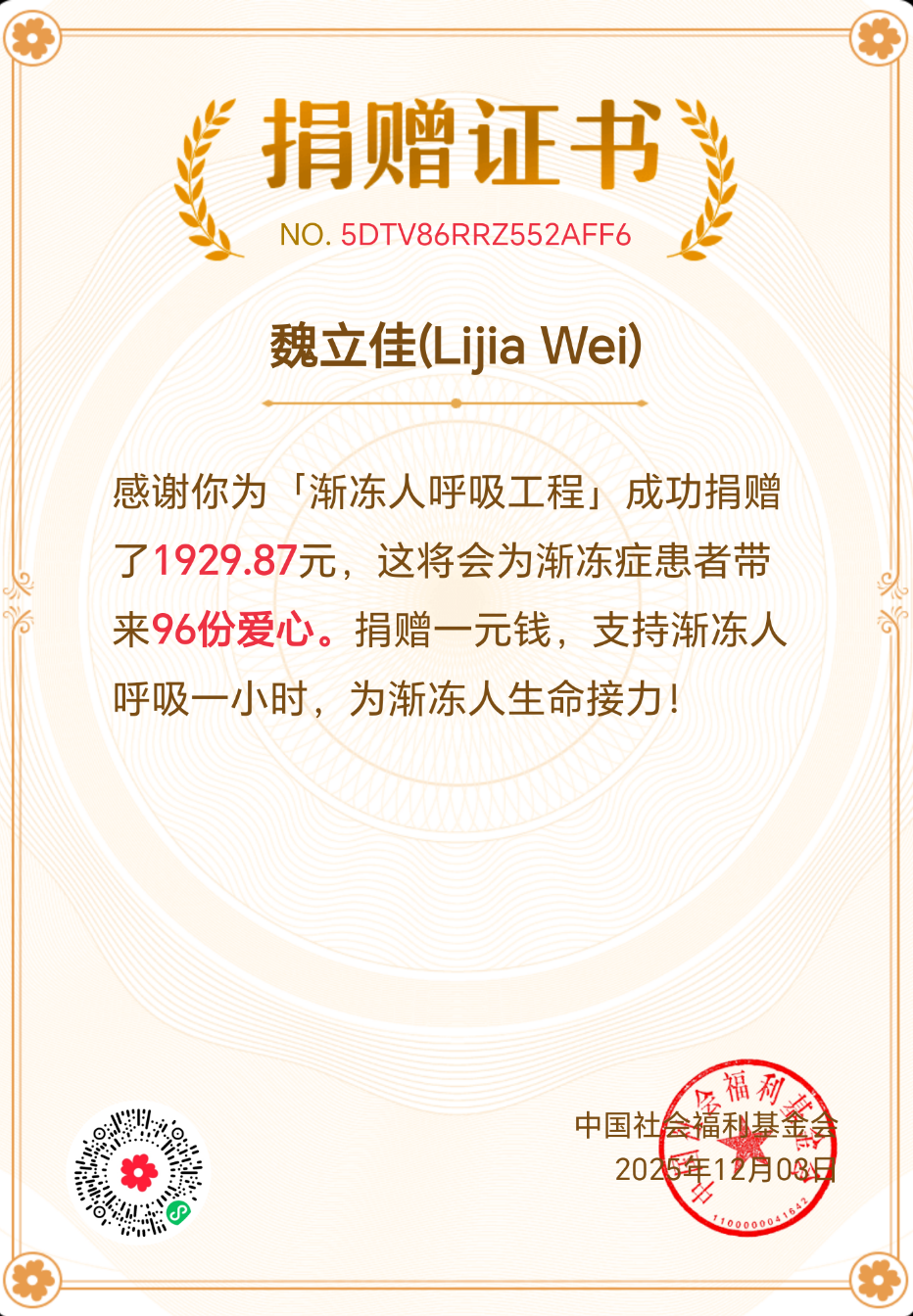}
    \caption{Certificate of the donation from main experiment}
    \label{fig:donation}
    \begin{flushleft}
        \footnotesize\textit{Notes:} The figure is the donation from the experiment, with a total donation of 1929.87 yuan to the Tencent Public Welfare.
    \end{flushleft}
\end{figure}

\end{appendices}
\end{document}